\documentclass[twocolumn]{aastex63}

\revised{\today}

\usepackage{graphicx}
\usepackage{bm}
\usepackage{chemformula}
\let\ce\ch
\usepackage{longtable}
\usepackage{supertabular}

\usepackage{textpos}
\usepackage{hyperref}
\usepackage{amsmath}
\usepackage{wasysym}

\newcommand{\beginappendixA}{%
        \setcounter{table}{0}
        \renewcommand{\thetable}{A\arabic{table}}%
        \setcounter{figure}{0}
        \renewcommand{\thefigure}{A\arabic{figure}}%
        \setcounter{equation}{0}
        \renewcommand{\theequation}{A\arabic{equation}}%
     }
     
 \newcommand{\beginappendixB}{%
        \setcounter{table}{0}
        \renewcommand{\thetable}{B\arabic{table}}%
        \setcounter{figure}{0}
        \renewcommand{\thefigure}{B\arabic{figure}}%
        \setcounter{equation}{0}
        \renewcommand{\theequation}{B\arabic{equation}}%
     }

\shorttitle{Towards RNA life on Early Earth}
\shortauthors{Pearce et al.}

\begin{document}

\title{Towards RNA life on Early Earth: From atmospheric HCN to biomolecule production in warm little ponds}



\author{Ben K. D. Pearce*}
\affiliation{Origins Institute and Department of Physics and Astronomy, McMaster University, 1280 Main St, Hamilton, ON, L8S 4M1, Canada}
\affiliation{Current address: Department of Earth and Planetary Science, Johns Hopkins University, Baltimore, MD, 21218, USA}
\thanks{Corresponding author: bpearce6@jhu.edu}

\author{Karan Molaverdikhani}
\affiliation{Landessternwarte, Zentrum f{\"u}r Astronomie der Universit{\"a}t Heidelberg, K{\"o}nigstuhl 12, 69117 Heidelberg, Germany}
\affiliation{Planet and Star Formation Department, Max Planck Institute for Astronomy, 69117 Heidelberg, Germany}
\affiliation{Universit{\"a}ts-Sternwarte, Ludwig-Maximilians-Universit{\"a}t M{\"u}nchen, Scheinerstrasse 1, D-81679 M{\"u}nchen, Germany}
\affiliation{Exzellenzcluster Origins, Boltzmannstraße 2, 85748 Garching, Germany}

\author{Ralph E. Pudritz}
\affiliation{Origins Institute and Department of Physics and Astronomy, McMaster University, 1280 Main St, Hamilton, ON, L8S 4M1, Canada}

\author{Thomas Henning}
\affiliation{Planet and Star Formation Department, Max Planck Institute for Astronomy, 69117 Heidelberg, Germany}

\author{Kaitlin E. Cerrillo}
\affiliation{Origins Institute and Department of Physics and Astronomy, McMaster University, 1280 Main St, Hamilton, ON, L8S 4M1, Canada}

\begin{abstract}
{\bf 
The origin of life on Earth involves the early appearance of an information-containing molecule such as RNA. The basic building blocks of RNA could have been delivered by carbon-rich meteorites, or produced in situ by processes beginning with the synthesis of hydrogen cyanide (HCN) in the early Earth's atmosphere. Here, we construct a robust physical and non-equilibrium chemical model of the early Earth atmosphere. The atmosphere is supplied with hydrogen from impact degassing of meteorites, sourced with water evaporated from the oceans, carbon dioxide from volcanoes, and methane from undersea hydrothermal vents, and in which lightning and external UV-driven chemistry produce HCN. This allows us to calculate the rain-out of HCN into warm little ponds (WLPs). We then use a comprehensive sources and sinks numerical model to compute the resulting abundances of nucleobases, ribose, and nucleotide precursors such as 2-aminooxazole resulting from aqueous and UV-driven chemistry within them. We find that at 4.4 bya (billion years ago) the limits of adenine concentrations in ponds for habitable surfaces is 0.05$\mu$M in the absence of seepage. These concentrations can be maintained for over 100 Myr. Meteorite delivery of adenine to WLPs can provide boosts in concentration by 2--3 orders of magnitude, but these boosts deplete within months by UV photodissociation, seepage, and hydrolysis. The early evolution of the atmosphere is dominated by the decrease of hydrogen due to falling impact rates and atmospheric escape, and the rise of oxygenated species such as OH from \ce{H2O} photolysis. Our work points to an early origin of RNA on Earth within $\sim$200 Myr of the Moon-forming impact.  
}
\end{abstract} 

\keywords{origin of life --- atmospheric chemistry --- early Earth --- astrobiology}

\section{Introduction}



Astrophysical, geophysical and fossil evidence suggests that life on Earth emerged in the interval of 4.5--3.7 bya (billion years ago) \citep{2018AsBio..18..343P}. A fundamental question about the origin of life such as our own is whether biomolecular building blocks critical to creating information polymers such as RNA and proteins can be synthesized {\it in situ} on a habitable planet \citep{Miller1953}. If not, then life's origin presumably depended on the delivery of biomolecules via external agents, such as carbon-rich meteorites \citep{1992Natur.355..125C,Reference425,Damer_Deamer2019}. 

HCN is a key biomolecule precursor because in aqueous solution it reacts with itself and other small molecules such as formaldehyde to produce several relevant biomolecules for the origin of RNA - widely thought to have been critical for the first life on Earth \citep{Rich1962,1986Natur.319..618G}. One advantage of HCN over other more complex biomolecule precursors is that there are multiple favorable reaction pathways for its production directly from the dissociation products of common atmospheric gases, i.e., \ce{N2}, \ce{CH4}, and \ce{H2} \citep{Pearce2020a}. The famous Miller-Urey experiments showed that reducing atmospheres rich in \ce{H2} and \ce{CH4} are favorable for HCN production, whereas oxidizing atmospheres rich in \ce{CO2} do not produce as much HCN \citep{Reference437,Schlesinger_Miller1983,Benner_et_al2020}. This is because oxygen must be removed from oxidized carbon (i.e. \ce{CO2} and \ce{CO}) before it can react to form HCN, which is energetically expensive. Reduced carbon (e.g. \ce{CH4}, \ce{CH3}), on the other hand, directly reacts to produce HCN \citep{Pearce2020b,Pearce2020a,2007AsNow..22e..76R}. It is the HCN produced by electrical discharges, once dissolved in the water reservoir in the Miller-Urey apparatus, that produces the plethora of amino acids \citep{Reference600,Reference440} and nucleobases \citep{Reference438,Becker_et_al2018}. 

However, such experiments do not address the whole planetary and geochemical context of an evolving planet and its atmosphere, nor do they address what conditions actually lead to RNA synthesis sufficient for an RNA world. Given the multiple processes that contribute to the balance of \ce{H2}, \ce{CH4}, and \ce{CO2} in the early Earth atmosphere, including volcanic outgassing, asteroid impacts, hydrothermal activity in undersea vents, hydrogen escape from the atmosphere, and rain-out, what are the yields of biomolecules in specific environments?  

Several invaluable observations are available to constrain early Earth conditions. The analysis of a zircon mineral inclusion has shown that the early mantle was already oxidized by $\sim$4.35 billion years ago (bya). This implies that by then, volcanoes mainly outgassed \ce{CO2} \citep{Reference119}. Before $4.35$ bya, isotopic evidence from the Earth's mantle (nitrogen, oxygen, titanium, calcium, chromium, nickel, ruthenium, molybdenum, neodymium, and deuterium) shows that accreting material was most similar to enstatite meteorites \citep{Dauphis_2017,Piani_et_al2020}. Reduced iron from these impactors would have been oxidized by water, releasing \ce{H2} \citep{Zahnle_et_al2020}. We note that recent isotopic studies suggest that substantial amounts of carbonaceous-chondrite-like materials were present in these impactors as well \citep{Fischer-Godde2020}. Past models predict that the early Earth atmosphere had a slightly reducing composition dominated by species such as \ce{N2}, \ce{CO2}, \ce{CH4}, \ce{CO}, and \ce{H2} \citep{Reference591,2011EPSL.308..417T,Zahnle_et_al2020}.

The ultimate step - actual RNA synthesis - could occur naturally in WLPs on the small land area available on the planet at that time \citep{Reference136}. The crucial point is that in the absence of any biological enzymes, bond formation that leads to RNA polymers involves thermal energy sufficient to remove water between the nucleotide building blocks. Such condensation reactions are well studied experimentally and arise naturally during seasonal or daily wet-dry cycles in WLPs \citep{Yi_et_al2020,Reference425,Ross_Deamer2016,Reference83,Morasch_et_al2014}. 

The route to nucleotides remains a big question in prebiotic chemistry. The older approach involved reacting nucleobases, ribose and a phosphorous source, which leads to low yields \citep{1963Natur.199..222P}. A more recent and radically different method bypasses the need for nucleobases and ribose reactants to obtain nucleotides, requiring simpler reactants of unknown concentration on early Earth such as glycolaldehyde, cyanamide, glyceraldehyde and cyanoacetylene \citep{Reference56}. The key intermediate in the latter pathway, sometimes referred to as the Powner--Sutherland approach, is 2-aminooxazole. 2-aminooxazole can also be formed by irradiating solutions of HCN to produce formaldehyde, which then reacts with another HCN molecule to produce glycolonitrile. Glycolonitrile then reacts further in the presence of aqueous HCN to produce glycolaldehyde and glyceraldehyde: two of the starting components of the Powner--Sutherland approach. Irradiating aqueous HCN in the presence of molar concentrations of NaCl and \ce{NH4Cl} salts produces cyanamide, another key reactant for this approach \citep{Yi_et_al2020}.

Another recent approach to nucleotide formation involves wet-dry cycling malononitrile and amidinium salts, the former of which could come from HCN-based reactions in WLPs \citep{Becker_et_al2018}. Finally, a route to the DNA nucleotides has been discovered by reacting nucleobases with acetaldehyde and the subsequent reaction with formaldehyde or glycolaldehyde \citep{Teichert_et_al2019}. Due to the uncertainty in the main prebiotic route to nucleotides, we take an agnostic approach and compute the WLP concentrations of nucleobases, ribose, and 2-aminooxazole produced via HCN aqueous chemistry driven by UV irradiation.

The ultimate source of HCN is from non-equilibrium photochemical and lightning-based reactions in the early Earth atmosphere. It has been modeled for a range of initial concentrations of key primordial species such as \ce{CH4} and \ce{CO2} \citep{Reference591,2011EPSL.308..417T}. Furthermore, experiments of electric discharges in reducing early Earth conditions have produced high abundances of HCN \citep{Ferus_et_al2017}. Most recently, rare, very large impactors ($\sim$400+ km) have been proposed as a source of enormous yields of HCN \citep{Zahnle_et_al2020,Benner_et_al2020,Reference439}. These early large impacts generate both the reducing conditions and the sustained high temperatures required to produce substantial \ce{CH4}: the main precursor to HCN. Very large impactors are also one logical explanation for the discrepancy between the highly-siderophile element (HSE) signatures in the lunar and terrestrial mantles \citep{Reference439}. However, impactors of such a large size are not necessarily a good solution for the origin of life. The energy from these impactors produces sustained high surface temperatures and pressures that render the planet uninhabitable (see Discussion). This would be a world devoid of ponds. Furthermore, the largest lunar impactor was $\sim$200 km in size---as seen in the cratering record---so it is expected that few objects greater than 300 km would have impacted early Earth \citep{Reference89}.

Here, we develop a holistic non-equilibrium atmosphere-pond coupled chemistry model that offers multiple important advancements to these past models. Instead of considering atmospheric processes individually, we build a comprehensive model that includes lightning, photochemistry, impact degassing, volcanism, ocean geochemistry and ocean evaporation simultaneously. We employ a newly extended disequilibrium atmospheric chemistry code that is augmented with new, expanded HCN and formaldehyde chemistry to follow atmospheric chemical evolution. These simulations are also informed by computed self-consistent pressure-temperature (P-T) profiles via radiative transfer.
The resulting computed HCN rain-out to surface WLPs drives the subsequent aqueous reactions of HCN into various critical biomolecules in the face of key terrestrial sinks such as hydrolysis, seepage, and UV photodestruction. The origin of formaldehyde (\ce{H2CO}) is also important to pin down as it is necessary for the formation of ribose, the pyrimidine nucleobases, and 2-aminooxazole.

\section{Methods}

\subsection{Early Earth Atmospheric Model}

\begin{figure*}[!hbtp]
\centering
\includegraphics[width=0.7\linewidth]{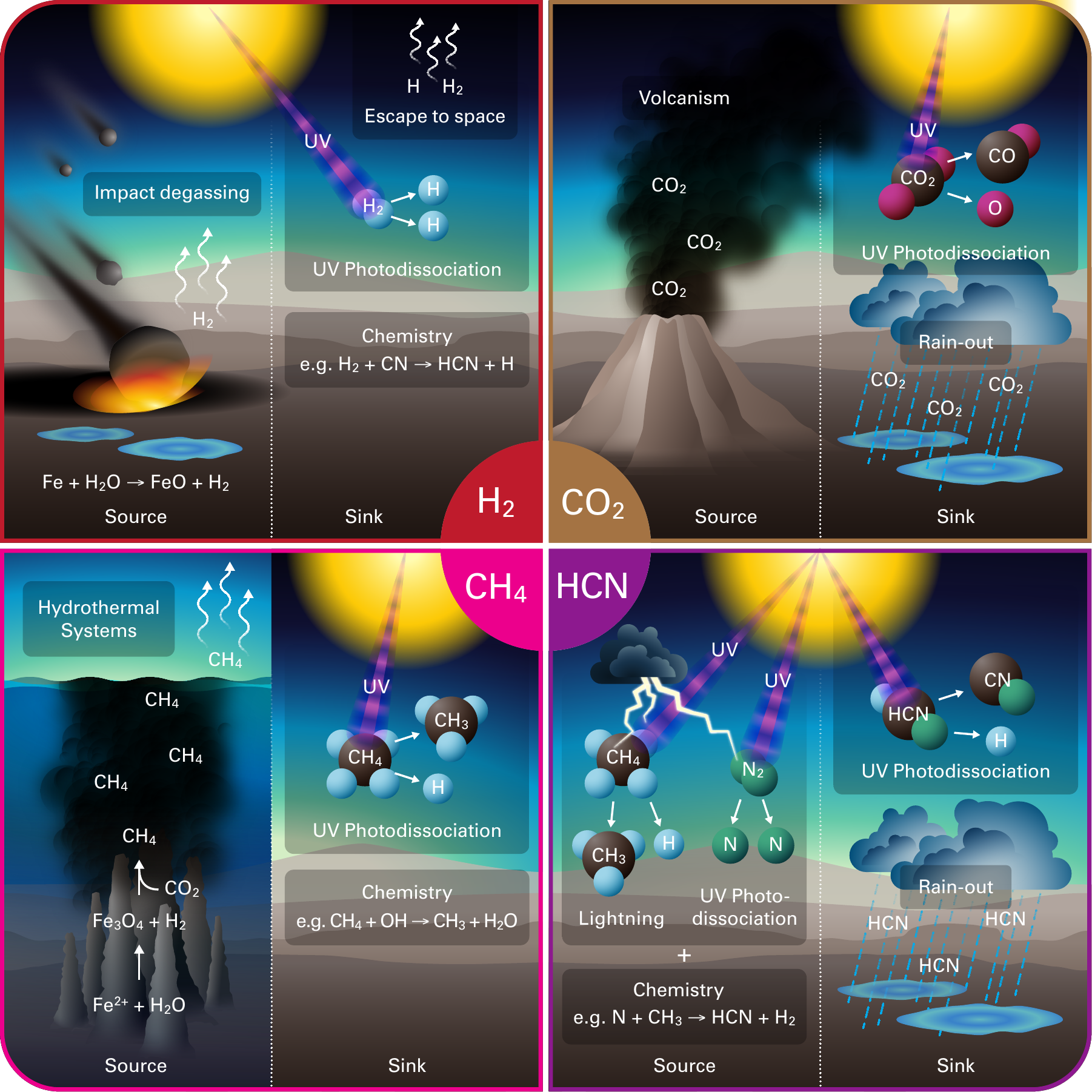}
\caption{An illustration of the sources and sinks of the four key species in our atmospheric model: \ce{H2}, \ce{CO2}, \ce{CH4}, and \ce{HCN}.}
\label{Fig1}
\end{figure*}

In Figure~\ref{Fig1}, we illustrate our early Earth atmospheric models by focusing on the main sources and sinks for the key molecular species relevant to controlling and determining HCN chemistry. As an overall principle, \ce{H2} and \ce{CO2} are the main species that determine whether the environment is reducing or oxidizing. It is the balance between these two species that determines the concentration of \ce{CH4} - the main precursor to HCN.

In the top left panel, impact degassing produces the \ce{H2}. The degassing rate at each epoch is calculated by combining equilibrium \ce{H2} production rates from enstatite chondrite impactors via the reaction \ce{Fe + H2O -> FeO + H2} \citep{Zahnle_et_al2020} with the bombardment rate on early Earth based on mathematical fits to the observed lunar cratering record \citep{Reference425,1990Natur.343..129C}. The main sinks for \ce{H2} include UV photodissociation, hydrodynamic escape to space, and disequilibrium chemistry.

In the top right panel, the main source for \ce{CO2} on early Earth is volcanic outgassing \citep{Zahnle_et_al2020}. We use a constant Earth-like volcanic \ce{CO2} outgassing rate in all our models \citep{Hu_et_al2012}. The main sinks for \ce{CO2} are photodissociation in the upper atmosphere, and rain-out in the lower atmosphere.


We utilize a potentially abundant source of \ce{CH4} to the Hadean atmosphere from serpentinization and Fischer-Tropsch-Type (FTT) synthesis. This begins with water-dependent processes in hydrothermal systems wherein Fe- and Mg-rich ultramafic rocks (e.g. olivine) in mid-ocean ridges and forearc systems produces \ce{H2}. Then, \ce{H2} reacts with the aqueous \ce{CO2} in these environments in the presence of mineral catalysts to produce \ce{CH4} \citep{Holm_et_al2015,Guzman-Marmolejo2013}. 
Abiotic methane production has been observed in hydrothermal systems \citep{Fiebig2007,Bradley2010}; however, experiments of FTT synthesis from olivine typically produce very low yields \citep{2018SciA....4.5747K}. Naturally occurring catalysts such as awaruite or chromite greatly speed up FTT synthesis \citep{Bradley2016}; however, the availability of these catalysts in primordial hydrothermal systems is still somewhat uncertain. Equilibrium models suggest this hydrothermal process can sustain $\sim$2--2.5 parts-per-million (ppm) of \ce{CH4} in the early atmosphere \citep{Guzman-Marmolejo2013}. We use the calculated \ce{CH4} outgassing rate from these models \citep{Guzman-Marmolejo2013}. The main sinks for methane are UV photodissociation and disequilibrium chemistry.

Finally, the main source of HCN is photodissociation or lightning dissociation of species such as \ce{N2} and \ce{CH4} followed by radical chemistry. In the case of lightning, this radical chemistry takes place at high temperature in the lightning channel. The main sinks for HCN are UV photodissociation and rain-out. For further details on the source and sink rates, see Appendix A.

We model the early Earth atmosphere during its reducing phase at 4.4 bya for calculated habitable surface temperatures of 78$^{\circ}$C (Model A) and 51$^{\circ}$C (Model C), as well as its oxidizing phase at 4.0 bya for calculated surface temperatures of 51$^{\circ}$C (Model B) and 27$^{\circ}$C (Model D). These particular surface temperatures result from the radiative transfer calculations for our chosen starting atmospheric compositions. These models differ in atmospheric composition, solar luminosity, UV irradiation intensity, HCN and radical production from lightning and impact bombardment rate (see Methods, Table~\ref{models}, for model details).

\subsection{Non-Equilibrium Atmospheric Chemistry}

ChemKM is a spatially 1D chemical kinetic model for disequilibrium atmospheric chemistry calculations that makes use of the Double precision Livermore Solver for Ordinary Differential Equations (DLSODE) from the ODEPACK collection \citep{Hindmarsh1982}. The error for this solver is controlled by the relative error tolerance and the absolute error tolerance, which are set to 10$^{-5}$ and 10$^{-99}$ respectively to insure numerical stability. 

ChemKM computes atmospheric chemistry based on input chemical network, P-T profile, eddy diffusion (mixing) profile, top-of-atmosphere (TOA) solar radiation, wavelength-dependent photochemical reactions, and influxes and outfluxes of species at the surface and TOA. ChemKM has been benchmarked with several other chemical kinetic codes\footnote{https://www.issibern.ch/teams/1dchemkinetics/}, and has been used in the past to simulate the atmospheres of Titan \citep{Pearce2020a}, as well as cold, hot and ultra-hot gaseous exoplanets \citep{Molaverdikhani_2019,Molaverdikhani2020}. Atmospheric rain-out was newly developed in ChemKM for this work. Simulations were run on the cluster for approximately 1 week. In this time, our models reached 1--50 million years of simulated time.

CRAHCN-O is a consistent reduced atmospheric hybrid chemical network now containing 259 two- and three-body reactions for the production of HCN and \ce{H2CO} in atmospheres dominated by any of \ce{H2}, \ce{CO2}, \ce{N2}, \ce{CH4}, and \ce{H2O}. We introduce 28 new reactions to CRAHCN-O in this work, in order to avoid the atmospheric build up of species that previously had no reaction sinks (see Tables~\ref{addedtwobodychemistry} and \ref{newthreebody} for details). We have tested an oxygenless version of this network (CRAHCN) by modeling HCN production in Titan's atmosphere, and our computed HCN profile agreed very well with the Cassini observations \citep{Pearce2020a}.

\subsection{Atmospheric Pressure-Temperature Profiles}

petitRADTRANS is a 1D radiative transfer code based on the correlated-k method for gas absorption and the Guillot temperature model \citep{2019AA...627A..67M,2010AA...520A..27G}. It is typically used to model exoplanet atmospheres to obtain  transmission and thermal spectra, e.g., \citet{Molaverdikhani_2019,2020AA...640A.131M,2020AJ....160..150W}. We build upon its existing functionality to calculate P-T profiles self-consistently with tropospheric water vapor in sequence with the Arden Buck equation \citep{Buck1981}.

P-T structure calculations are performed using petitRADTRANS \citep{2019AA...627A..67M}, and atmospheric chemistry is calculated using ChemKM \citep{Molaverdikhani_2019,Molaverdikhani2020} coupled with an updated version of the CRAHCN-O chemical network \citep{Pearce2020b,Pearce2020a}. We are the first to calculate composition-dependent P-T profiles for modeling HCN chemistry in the early Earth atmosphere. Past models have used a general habitable P-T profile, or estimated the surface temperature using an analytic equation for a moist adiabat \citep{Reference591,2011EPSL.308..417T,Zahnle_et_al2020}. The code used to compute P-T profiles is available at https://gitlab.com/mauricemolli/petitRADTRANS.
 
For complete details on our atmospheric models, see the Appendix A.

\subsection{Lightning Production of Molecules}

We follow the thermodynamic treatment from \citet{Chameides_Walker1981} for the lightning production of HCN and other species on early Earth. Based on our initial atmospheric compositions, we calculate the equilibrium abundances of \ce{HCN}, \ce{H2}, \ce{N2}, \ce{H2O}, \ce{CO2}, \ce{CH4}, \ce{O2}, \ce{NO}, \ce{OH}, \ce{H}, \ce{^4N}, \ce{CO}, \ce{^3O}, and \ce{CH3}, in a 1 cm$^2$ lightning channel extending through the lowest layer in our atmospheres at a freeze out temperature T$_F$ = 2000 K. We use thermochemical data from the JANAF tables \citep{Stull_Prophet1971}, and the ChemApp Software library for Gibbs free energy minimization (distributed by GTT Technologies, http://gtt.mch.rwth-aachen.de/gtt-web/).

We then use the resultant mixing ratios to calculate the influx of each of these species into the lowest layer of our atmospheres. These species were chosen as they are dominant equilibrium products in the early Earth lightning models by \citet{Chameides_Walker1981}. The freeze out temperature (T$_F$) was chosen to most accurately model the HCN produced in a lightning strike, as this is the key species of interest in this paper. Although freeze out temperatures typically range from 1000--5000 K across species, one freeze out temperature must be chosen to conserve the elemental abundances in the lightning strike. A non-equilibrium approach was also considered; however, an extensive high-temperature (up to 30000 K) chemical kinetic network would be required and is perhaps unnecessary given the $< \mu$s equilibrium timescale above 10000 K compared to the $\sim$10$\mu$s timescale of eddy diffusion, and the $\sim$100 ms cooling time of a lightning channel \citep{Hill_et_al1980}.


\begin{table*}[ht!]
\centering
\caption{Summary of the four early Earth atmospheric models in this work. Initial compositions for each model are chosen to A) align with typical assumptions for reducing (\ce{H2}-dominant) or oxidizing (\ce{CO2}-dominant) conditions at the chosen epoch, and B) yield calculated P-T profiles with habitable surface conditions (i.e. 0 $^{\circ}$C $\leq$ T$_s$ $\leq$ 100 $^{\circ}$C) (see Figure~\ref{P-T} for P-T profiles).\label{models}} 
\begin{tabular}{ccccllll}
\\
\multicolumn{1}{l}{Model} &  
\multicolumn{1}{l}{Description} & 
\multicolumn{1}{l}{Date (bya)} &  
\multicolumn{1}{l}{P$_s$ (bar)} & 
\multicolumn{1}{l}{T$_s$ ($^{\circ}$C)} & 
\multicolumn{1}{l}{Molar Composition} & 
\multicolumn{1}{l}{Surface flux ($\frac{1}{cm^{2}s}$)} &
\multicolumn{1}{l}{Lightning ($\frac{1}{cm^{2}s}$)}
\\[+2mm] \hline \\[-2mm]
A & Early Hadean & 4.4 & 1.5 & 78 & \ce{H2}: 90\% & \ce{H2}: 2.3$\times$10$^{11}$ & \ce{H}: 4.1$\times$10$^{6}$ \\ 
& (Reducing) & & & & \ce{N2}: 10\% & \ce{CO2}: 3.0$\times$10$^{11}$ & \ce{CO}: 6.3$\times$10$^{3}$\\ 
& & & & & \ce{CH4}: 2 ppm  & \ce{CH4}: 6.8$\times$10$^{8}$ & \ce{OH}: 4.4$\times$10$^{3}$ \\
& & & & & \ce{H2O}: Figure~\ref{water_profiles} & \ce{H2O}: 2.0$\times$10$^{9}$ & \ce{NO}: 5.0$\times$10$^{1}$\\
& & & & & & & \ce{^3O}: 4.4$\times$10$^{0}$ \\
& & & & & & & \ce{^4N}: 7.2$\times$10$^{-1}$ \\
& & & & & & & \ce{HCN}: 4.4$\times$10$^{-1}$ \\
& & & & & & & \ce{O2}: 2.2$\times$10$^{-2}$ \\
& & & & & & & \ce{CH3}: 1.6$\times$10$^{-3}$ \\
\\[-2mm] \hline \\[-2mm]
B & Late Hadean & 4.0 & 2 & 51 & \ce{CO2}: 90\% & \ce{H2}: 2.3$\times$10$^{10}$ & \ce{CO}: 1.8$\times$10$^{5}$ \\
& (Oxidizing) & & & & \ce{N2}: 10\% & \ce{CO2}: 3.0$\times$10$^{11}$ & \ce{O2}: 8.5$\times$10$^{4}$ \\
& & & & & \ce{CH4}: 10 ppm & \ce{CH4}: 6.8$\times$10$^{8}$ & \ce{NO}: 7.4$\times$10$^{3}$ \\
& & & & & \ce{H2O}: Figure~\ref{water_profiles} & \ce{H2O}: 2.0$\times$10$^{9}$ & \ce{OH}: 3.5$\times$10$^{3}$ \\
& & & & & & &  \ce{^3O}: 5.5$\times$10$^{2}$ \\
& & & & & & & \ce{H}: 9.7$\times$10$^{1}$ \\
& & & & & & & \ce{^4N}: 3.2$\times$10$^{-3}$ \\
& & & & & & & \ce{HCN}: 2.9$\times$10$^{-6}$ \\
\\[-2mm] \hline \\[-2mm]
C & Early Hadean & 4.4 & 1.13 & 51 & \ce{H2}: 90\% & \ce{H2}: 2.3$\times$10$^{11}$ & \ce{H}: 1.1$\times$10$^{5}$ \\ 
& (Reducing) & & & & \ce{N2}: 10\% & \ce{CO2}: 3.0$\times$10$^{11}$ & \ce{OH}: 1.3$\times$10$^{2}$ \\ 
& & & & & \ce{CH4}: 1 ppm  & \ce{CH4}: 6.8$\times$10$^{8}$ & \ce{CO}: 8.1$\times$10$^{1}$ \\
& & & & & \ce{H2O}: Figure~\ref{water_profiles} & \ce{H2O}: 2.0$\times$10$^{9}$ & \ce{NO}: 1.6$\times$10$^{0}$\\
& & & & & & & \ce{^3O}: 1.6$\times$10$^{-1}$ \\
& & & & & & &  \ce{^4N}: 2.2$\times$10$^{-2}$ \\
& & & & & & & \ce{HCN}: 4.1$\times$10$^{-3}$ \\
& & & & & & & \ce{O2}: 7.4$\times$10$^{-4}$ \\
& & & & & & & \ce{CH3}: 1.3$\times$10$^{-5}$ \\
\\[-2mm] \hline \\[-2mm]
D & Late Hadean & 4.0 & 2 & 27 & \ce{CO2}: 90\% & \ce{H2}: 2.3$\times$10$^{10}$ & \ce{CO}: 1.8$\times$10$^{5}$ \\
& (Oxidizing) & & & & \ce{N2}: 10\% & \ce{CO2}: 3.0$\times$10$^{11}$ & \ce{O2}: 8.5$\times$10$^{4}$ \\
& & & & & \ce{CH4}: 1.5 ppm & \ce{CH4}: 6.8$\times$10$^{8}$ & \ce{NO}: 7.4$\times$10$^{3}$ \\
& & & & & \ce{H2O}: Figure~\ref{water_profiles} & \ce{H2O}: 2.0$\times$10$^{9}$ & \ce{OH}: 3.5$\times$10$^{3}$ \\
& & & & & & &  \ce{^3O}: 5.5$\times$10$^{2}$ \\
& & & & & & & \ce{H}: 9.7$\times$10$^{1}$ \\
& & & & & & & \ce{^4N}: 3.2$\times$10$^{-3}$ \\
& & & & & & & \ce{HCN}: 2.9$\times$10$^{-6}$ \\
\\[-2mm] \hline
\multicolumn{8}{l}{\footnotesize P$_s$: Surface Pressure} \\
\multicolumn{8}{l}{\footnotesize T$_s$: Surface Temperature} \\
\end{tabular}
\end{table*}


\subsection{Complete Impact-Atmosphere-Ocean Coupling Models}

The main assumption of our models is that the surface of the Earth maintained habitability (i.e. 0$^{\circ}$C $<$ T $<$ 100$^{\circ}$C), which is key for the presence of WLPs and the origin of life. We begin with assumed reducing and oxidizing atmospheric compositions for the early and late Hadean, respectfully, and calculate the initial P-T profiles and tropospheric water vapor based on these compositions. We adjust both initial methane concentration and surface pressure to obtain calculated temperature profiles that fall within the habitable range. We smooth the initial water profiles from our calculations to 1\% at the surface, and include ocean-atmosphere coupling by imposing an ocean evaporation rate of 2$\times$10$^{9}$ cm$^{-2}$ s$^{-1}$ to maintain a water mixing ration of $\sim$0.1--1\% at the surface. 

In Table~\ref{models}, we summarize the four early Earth atmospheric models in our study. We model two epochs which vary in atmospheric composition, solar luminosity, UV irradiation intensity, and asteroid bombardment rate. These models correspond to the early Hadean, at 4.4 bya (billion years ago) and the late Hadean, at 4.0 bya. We compute two habitable P-T profiles for each model by slightly adjusting the methane content and/or surface pressure. The luminosity, UV intensity, and asteroid bombardment rate at each epoch are based on stellar evolution models \citep{Heller2020,2015AA...577A..42B}, observations of solar analogs \citep{Ribas_et_al2005}, and the lunar cratering record \citep{1990Natur.343..129C}, respectively. 


\subsection{Biomolecule Chemistry in Warm Little Ponds: Sources and Sinks}

Our atmospheric models are coupled (via rain-out) with the sources and sinks warm little pond model (WLP) we first developed in \citet{Reference425}. Biomolecule abundances are described by first-order linear differential equations and are solved numerically. The evolving concentrations of nucleobases, ribose, formaldehyde, and 2-aminooxazole in our WLP models are driven by the rate of incoming HCN from rain-out, and biomolecule losses due to UV dissociation, seepage, and hydrolysis. Given experimental reaction rates are fast ($\lesssim$ days), we apply experimental reaction yields to our HCN pond concentrations in order to estimate the pond concentrations of formaldehyde, nucleobases, ribose, and 2-aminooxazole. However, we recognize that at the lower pond concentrations here, these reactions could take much longer. The code used to compute biomolecule concentrations in ponds is available at https://github.com/bennski/Wet\_Dry\_Cycling\_Pond\_Model.

We utilize the experimental result that ultraviolet (UV) irradiation of liquid water produces solvated electrons, enabling a chemical pathway from HCN to formaldehyde (\ce{H2CO}) \citep{Yi_et_al2020} in ponds. \ce{H2CO} can also enter ponds directly from the atmosphere \citep{Pinto_et_al1980}. Aqueous solutions containing HCN and \ce{H2CO} can produce nucleobases (i.e., adenine, guanine, cytosine, uracil, thymine) \citep{1961Natur.191.1193O,2008OLEB...38..383L,2019AA...626A..52F}, which are the base-pairing components of RNA and DNA, as well was ribose \citep{Butlerow1861,Breslow1959}, which binds with phosphate to make up the RNA backbone. 

Furthermore, irradiated and wet-dry cycled or flowing solutions of HCN in the presence of phosphorous and dissolved salts enable the production of 2-aminooxazole: a key intermediate in the Powner--Sutherland pathway for producing the pyrimidine building blocks of RNA (cytidine and uridine monophosphate) \citep{Yi_et_al2020,Ritson_et_al2018,Reference56}. Finally, \citet{Becker_et_al2018} recently presented a pathway to RNA nucleosides that involves the wet-dry cycling of solutions containing HCN and other atmospheric precursors. Such prebiotic chemistry experiments and models are based on the assumption that species such as HCN and \ce{H2CO} would be present and concentrated in WLPs on early Earth. 

We considered including formose-like reactions occurring from more complex aldehydes (e.g. glycolaldehyde, glyceraldehyde) \citep{Cassone_et_al2018}; however, given the greater complexity of these species, it is expected that they would be produced in the atmosphere in much lower concentrations than HCN and \ce{H2CO}. However, it would be valuable to explore reactions that produce these more complex aldehydes in planetary atmospheres so that these sugar precursors can be included in a future model.

Lastly, we understand that straightforwardly applying experimental biomolecule production yields from HCN to obtain estimates of biomolecule influxes in our WLP model has issues given the high reactant concentrations and ideal conditions of each experimental setup; however, given that there are no chemical kinetic rate coefficients for the aqueous production of nucleobases, ribose and 2-aminooxazole, we utilize experimental yields in our model to obtain reasonable first order estimates. We emphasize that these biomolecule concentrations should be understood as upper bounds in the absence of any concentrating mechanism beyond evaporation.

In Table~\ref{biomolecule_yields} we summarize the sources and sinks of our pond models. See Section~\ref{methods2} and \citet{Reference425} for complete details regarding these models.

\subsection{Warm-Wet Cycling in WLPs}

We have found that ponds that are roughly 1 meter in radius and depth and are cylindrical in shape are an optimal fiducial estimate for subsequent RNA polymer synthesis by wet-dry cycles \citep{Reference425}. We use the ``intermediate'' hot early Earth environment from \citet{Reference425}, which is based on the seasonal sinusoidal precipitation rates in Indonesia \citep{Reference113,Reference117}. Precipitation coupled with evaporation and seepage produces a natural wet-dry cycle within the pond that has a $\sim$6-month wet phase followed by a $\sim$6-month dry phase. Various pond environments were explored in \citet{Reference425}, and were found to produce similar results in terms of peak nucleobase concentrations.

\section{Results}

\subsection{Atmospheric \ce{HCN} and \ce{H2CO}}

\begin{figure*}[!hbtp]
\centering
\includegraphics[width=0.9\linewidth]{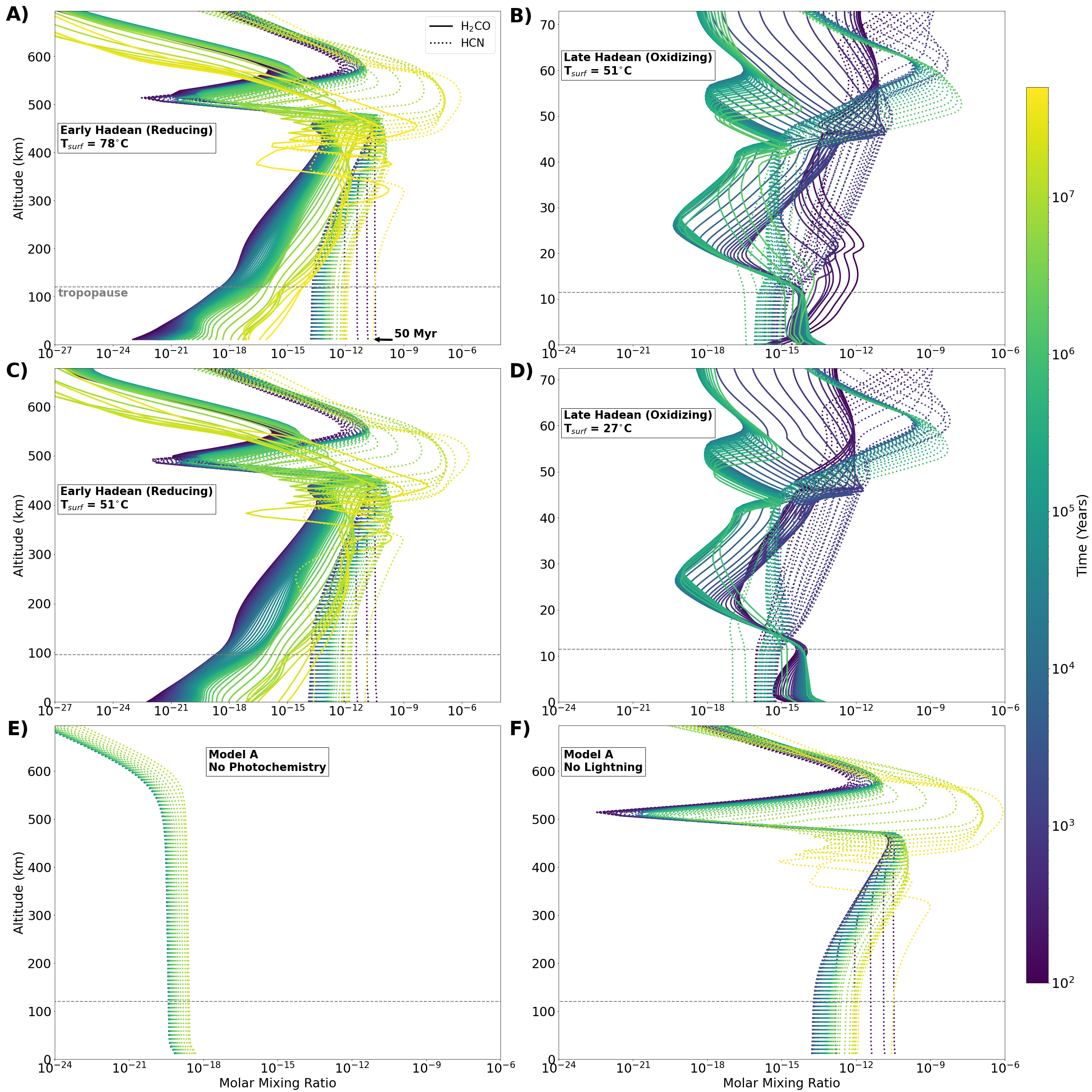}
\caption{{\bf A--D)} \ce{H2CO} and HCN atmospheric mixing ratios from $t$ = 100 years to 1--50 million years for the four early Earth models listed in Table~\ref{models}. Pressures go from 1.13--2 bar at the surface to 10$^{-8}$ bar at the top of atmosphere. Atmospheric scale heights vary primarily due to differences in mean molecular weight. The tropopause is labeled and corresponds to a pressure of $\sim$0.14 bar. {\bf E)} HCN atmospheric mixing ratio from $t$ = 100 years to 13 million years for model A with photochemistry turned off. {\bf F)} HCN atmospheric mixing ratio from $t$ = 100 years to 50 million years for model A with lightning chemistry turned off.}
\label{HCN_H2CO_plot}
\end{figure*}

In Figure~\ref{HCN_H2CO_plot}A, B, C, and D, we plot the temporal evolution of atmospheric HCN and \ce{H2CO} in these four early Earth models. To give some context for what would be considered high atmospheric HCN abundances, Cassini observed HCN mixing ratios in the heavily reducing atmosphere of Titan to be $\sim$10$^{-7}$--10$^{-6}$ near the surface (150--300 km), and $\sim$10$^{-4}$--10$^{-2}$ in the upper atmosphere (700--1050 km) \citep{2010Icar..205..559V,2011Icar..214..584A,2011Icar..216..507K,2009PSS...57.1895M}. 

In the early Hadean (reducing) model A, HCN mixing ratios decrease from 10$^{-11}$ at the surface to 10$^{-14}$ from $t$ = 100 years to 400 years. Then, HCN slowly builds back up again to 2.7$\times$10$^{-11}$ over the next 50 million years. Moving up in altitude we see the main region of HCN production from UV radiation at $\sim$500--600 km, which provides an HCN abundance of $\sim$10$^{-7}$--10$^{-6}$ at these altitudes after 10 million years. We isolate the predominant energy source for HCN production in Figures~\ref{HCN_H2CO_plot}E and F by turning off UV photochemistry and lightning chemistry, respectively, for our early Hadean (reducing) model A.

In the late Hadean (oxidizing) model B, HCN mixing ratios decrease at the surface from 10$^{-14}$ to 10$^{-16}$ from $t$ = 100 years to $\sim$40,000 years. Then, HCN slowly builds back up to 10$^{-15}$ over the next 2 million years. UV production of HCN produces a peak abundance of $\sim$10$^{-8}$ at $\sim$50 km at 2 million years.


One of the most important results is that neither oxidizing model produces nearly as much HCN as our reducing models. Our calculations reveal that HCN production near the surface is about 4 orders of magnitude more favorable in reducing conditions than it is in oxidizing conditions. It is worth noting that the oxidizing conditions at 4.0 bya have reduced gases such as \ce{H2} in the ppb--ppm range. Even more oxidizing conditions could be present after 4.0 bya with the declining \ce{H2} impact degassing rate allowing oxygen species to become more dominant.


We learned that UV photochemistry is crucial for atmospheric HCN production on early Earth (Figures~\ref{HCN_H2CO_plot}E and F). A major result of our simulations is that without UV photochemistry, HCN would be over 8 orders of magnitude less abundant at the surface during the early Hadean. Even with an increased lightning flash density representative of storms occurring during volcanic eruptions (1$\times$10$^4$ flashes km$^{-2}$ yr$^{-1}$  \citep{Hodosan2016}), we see no enhancement in HCN concentration beyond the photochemical result (see Figure~\ref{maxlightning_plot}).

Formaldehyde, given these atmospheric models, would need to come from elsewhere, such as UV driven, aqueous chemistry in WLPs. In comparison to HCN, \ce{H2CO} is much less abundant at the surface in our early Hadean (reducing) models. \ce{H2CO} builds up from $\sim$10$^{-23}$ to $\sim$10$^{-16}$ over 50 million years. The mixing ratio for \ce{H2CO} is at its highest value of 10$^{-9}$--10$^{-8}$ in the mid atmosphere of these models after 50 million years. In the late Hadean (oxidizing) model B, \ce{H2CO} increases at the surface from $\sim$10$^{-16}$ to $\sim$10$^{-13}$ in $\sim$1 million years. We did not explore atmospheric \ce{H2CO} production further given its considerably low abundances in all models.

\subsection{Surface Abundances}

\begin{figure*}[!hbtp]
\centering
\includegraphics[width=\linewidth]{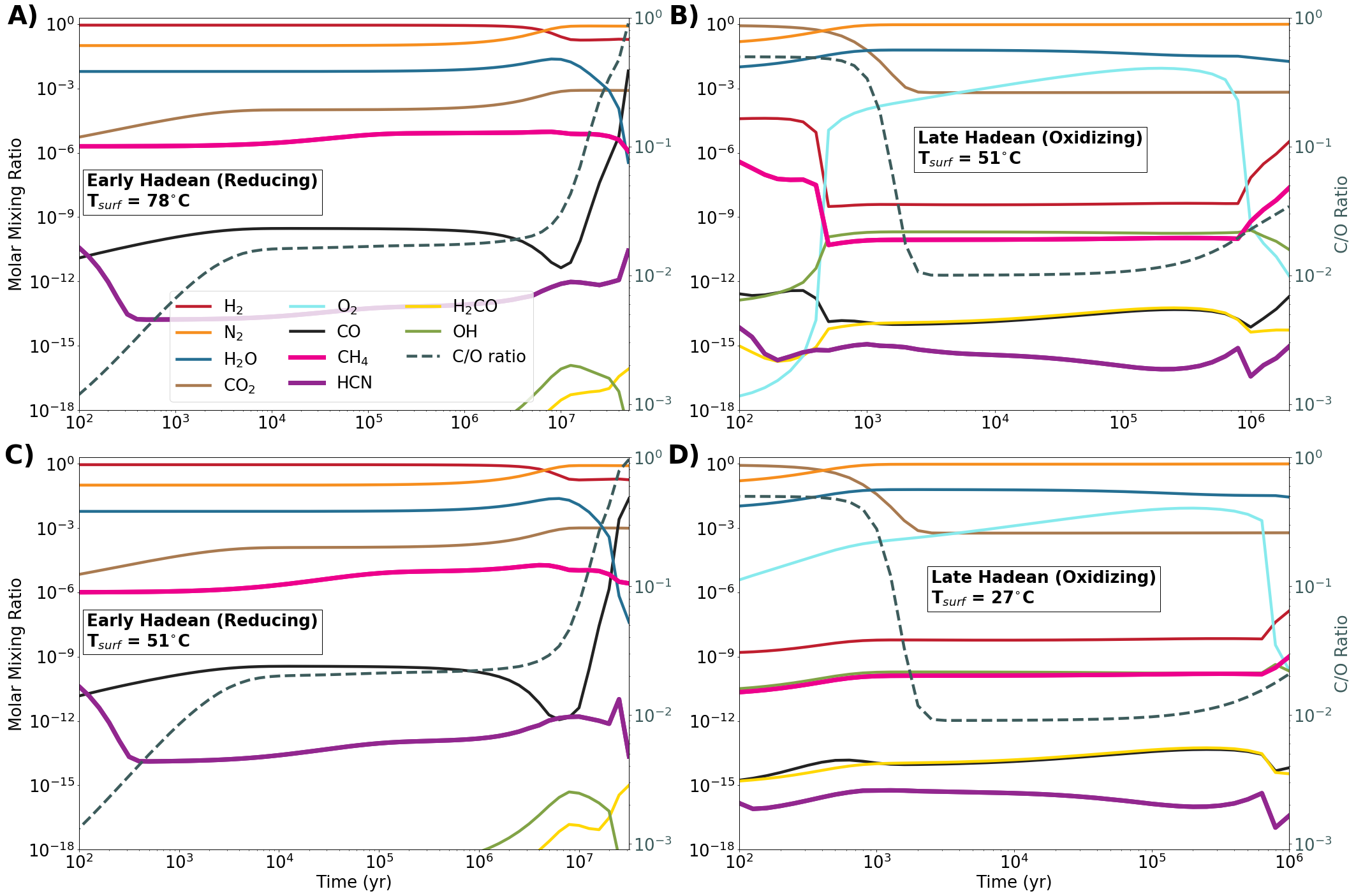}
\caption{Abundances of key species in the lowest atmospheric layer as a function of time in our four early Earth models. The evolving C/O ratio at the surface layer is the dotted line with values labeled on the right side of the y-axis. Model details are in Table~\ref{models}.}
\label{Surf_Abund_plot}
\end{figure*}

The temporal evolution of the dominant atmospheric species at the lowest (surface) layer in the atmosphere is shown in Figure~\ref{Surf_Abund_plot}. The atmospheric rain-out rates for HCN and \ce{H2CO} from this layer provides the influx rates into the WLPs.

One striking result of our atmospheric chemistry calculations is that the HCN and \ce{CH4} mixing ratios are correlated after the first few hundred years in all our models. The relevant abundance curves are bolded in Figure~\ref{Surf_Abund_plot} to emphasize this point, as the HCN and \ce{CH4} abundances trace one another closely from $t$ = 300 years onward.

In Figure~\ref{HCN_CH4_plot}, we plot the molar ratio of HCN to \ce{CH4} over time and find an average value of $\sim$1--4$\times$10$^{-7}$ for the early Hadean (reducing) models and 3--5$\times$10$^{-6}$ for the late Hadean (oxidizing) models. 


\ce{CH4} and \ce{HCN} surface abundances also follow the trend of the \ce{H2} surface abundances in the late Hadean (oxidizing) models B and D. This is because \ce{H2} drives the evolution of reducing and oxidizing conditions in our models. We see that \ce{CH4} abundances are stable in reducing (high \ce{H2}) conditions, and that \ce{CH4} abundances are depleted in oxidizing (high \ce{O2} and \ce{OH}) conditions.

There is an anti-correlation between \ce{HCN} and \ce{O2} in the late Hadean (oxidizing) models. This is because a high \ce{O2} environment leads to the oxidation of carbon into species such as \ce{H2CO} rather than the reduction of carbon into HCN. This is also why we also see a correlation between \ce{O2} and \ce{H2CO} in the late Hadean (oxidizing) models.

These correlations and anti-correlations are consistent with the atmospheric observations we see for Titan and present-day Earth, respectively. Titan's atmosphere is abundant in HCN (10$^{-7}$--10$^{-2}$) \citep{2010Icar..205..559V,2011Icar..214..584A,2011Icar..216..507K,2009PSS...57.1895M} due to the high abundances of reducing gases such as \ce{CH4} ($\sim$5.7\%), and \ce{H2} ($\sim$0.1\%) and low concentrations of oxidizing gases such as \ce{CO2} (10--20 ppb) and \ce{H2O} (0.5--8 ppb) \citep{Reference143,2017JGRE..122..432H}. On the other hand, HCN on Earth today is present in low abundances $\sim$10$^{-10}$ \citep{Cicerone_Zellner1983} because of the high abundance of oxidizing gases in our atmosphere such as \ce{O2} (21\%), \ce{H2O} (0--3\%) and \ce{CO2} ($\sim$400 ppm) and modest $\sim$1 ppm levels of \ce{CH4}.


We tested the hypothesis presented in several exo-atmosphere studies that the C/O ratio plays a central role in controlling their chemical composition \citep{2012ApJ...758...36M,2015ApJ...813...47M,2019Icar..329..124R,Molaverdikhani_2019,2019ApJ...873...32M,2020ApJ...899...53M}. In our early Hadean (reducing) models, the C/O ratio at the lowest atmospheric layer increases from 10$^{-3}$ to 10$^{-2}$ in the first few thousand years, and then to 1 after $\sim$30 million years. In the late Hadean (oxidizing) models, the C/O ratio decreases from 0.5 to $\sim$10$^{-2}$ in the first few thousand years, and then increases to 0.02--0.03 over the next 1--2 million years. Evidently the C/O ratios in our models are set mainly by the increase and decrease of \ce{CO2} and \ce{CO} at the surface, and in the late Hadean (oxidizing) models, also by the fluctuation of \ce{O2}. Our models show that the abundances of key biomolecule precursor species \ce{CH4} and HCN are not strongly dependent on the C/O ratio.

These results are not entirely surprising. In a study of the effects of C/O ratio on atmospheric HCN abundance, \citet{2019Icar..329..124R} found that for a given C/O ratio, the presence of methane leads to considerably more HCN. Considering these results, we suggest that \ce{CH4} abundance is a better fundamental driver for HCN production than the C/O ratio on its own.

\subsection{Biomolecule Concentrations in WLPs}

\begin{figure*}[!hbtp]
\centering
\includegraphics[width=\linewidth]{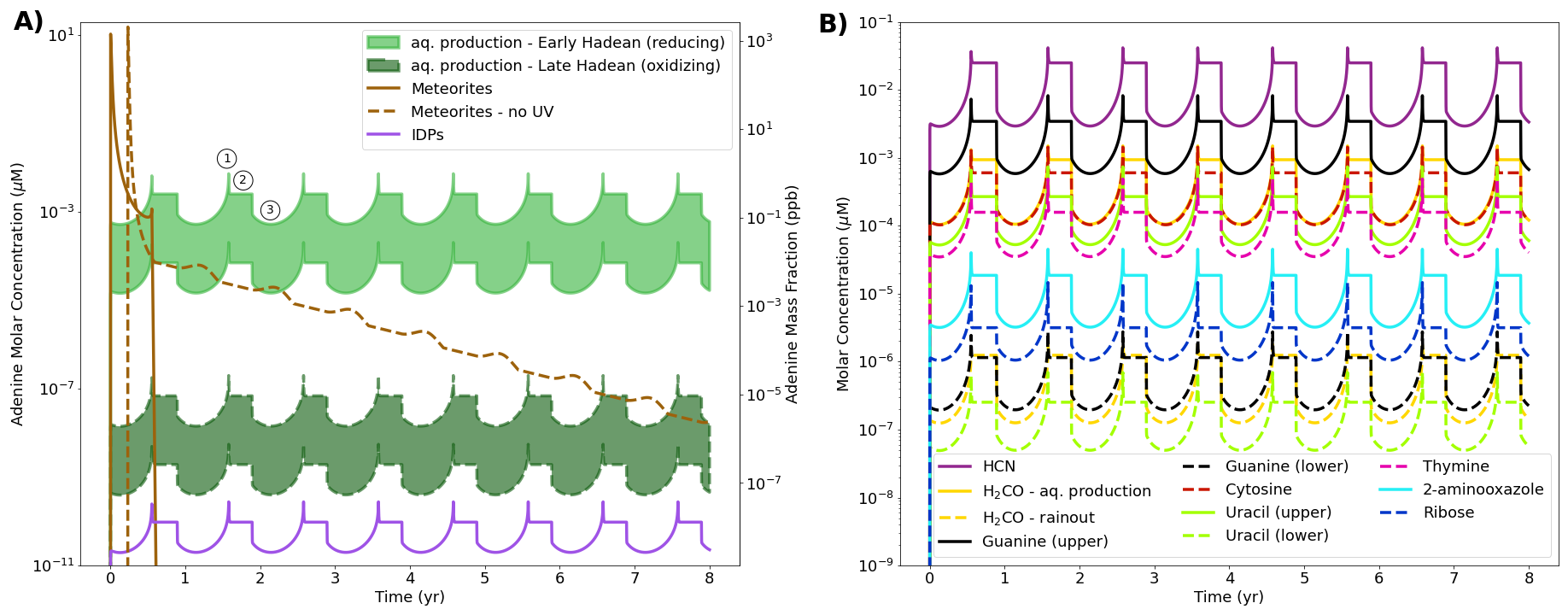}
\caption{{\bf A)} Comparative histories of adenine concentrations in warm little ponds from aqueous production (this work), versus delivery from meteorites and interplanetary dust particles (IDPs) \citep{Reference425}. Concentrations are calculated using the sources and sinks pond model developed in \citet{Reference425} that cycles between $\sim$6 months of wet and $\sim$6 months of dry conditions; the one exception is the ``Meteorites - no UV'' model which is calculated for a pond that never dries up and for which UV is never turned on. Aqueous production of adenine is sourced from atmospheric rain-out of HCN multiplied by the range of experimental yields (see Table~\ref{biomolecule_yields}). Sinks include UV photodissociation in the dry phase, and hydrolysis and seepage in the wet phase. At location (1), the pond has dried down to 1 mm creating the maximum concentration. At location (2), UV irradiation is turned on and the concentration reduces until production from HCN influx and destruction from UV dissociation equilibrate. Finally, at location (3), precipitation has filled the pond up to its highest point, resulting in the concentration minimum. This cycle repeats annually. Hydrolysis has no affect on these curves, as the other two sinks are more efficient and occur on shorter timescales. The meteorite and IDP curves are taken directly from \citet{Reference425}. {\bf B)} Calculations of pond concentrations of various biomolecules as a result of atmospheric rain-out of HCN or \ce{H2CO} (yellow dotted line) for our early Hadean (reducing) model A. See Table~\ref{biomolecule_yields} for the experimental yields and sink rates used in the model calculations.}
\label{Pond_Models}
\end{figure*}

In Figure~\ref{Pond_Models}A, we display the concentrations of adenine in our model WLP from aqueous production for different HCN rain-out (influx) rates from our early Hadean (reducing) and late Hadean (oxidizing) atmospheric models (see Figure~\ref{rain-out_plot} for rainout rates). Adenine concentrations are displayed as shaded regions to cover the range of experimental yields of adenine production from HCN. We also display for comparison the concentrations of adenine from meteoritic and interplanetary dust particle (IDP) delivery calculated using the same source/sink pond models in \citet{Reference425}.

Given the range of experimental conditions, there are additional uncertainties beyond the shaded regions that are not displayed here. Aqueously produced biomolecule concentrations could be lower than the abundances calculated here due to 1) the low nM-range HCN pond concentrations compared to typically molar-range experimental concentrations, 2) differences in the radiation field used in experiments for photolytic \ce{H2CO} production versus the solar radiation incident on WLPs, and 3) uncertainties in the hydrolysis and photodestruction rates for each biomolecule. Conversely, biomolecule concentrations could be higher than our calculated values due to various concentration mechanisms including adsorption to mineral surfaces, and sequestration into mineral gels or amphiphilic matrices \citep{Damer_Deamer2019,Deamer2017}. Given these uncertainties, we present the following biomolecule concentrations as an upper bound in the absence of concentration mechanisms.


Adenine concentrations from aqueous production peak at 7.3 nM (1 ppb) and 0.2 pM (0.03 pptr) for our early Hadean (reducing) and late Hadean (oxidizing) models, respectively. The early Hadean (reducing) adenine concentrations are approximately 3 orders of magnitude smaller than the peak adenine concentration from meteoritic delivery of 10.6 $\mu$M (1.43 ppm); however, the adenine concentration from aqueous production can be sustained for more than 100 million years rather than days.


Adenine concentrations from delivery by IDPs are the most dilute in our WLP models, peaking at $\sim$10$^{-10}$ $\mu$M \citep{Reference425}.

In Figure~\ref{Pond_Models}B, we plot the pond concentrations of HCN and \ce{H2CO} from atmospheric rain-out, as well as the concentrations of nucleobases, ribose, \ce{H2CO} and 2-aminooxazole from aqueous HCN-based production. HCN concentrations peak at 0.04 $\mu$M and reduce to approximately 3 nM when the water level in the pond is highest.

We learn from these models that formaldehyde in WLPs likely did not come directly from the atmosphere during the early Hadean. \ce{H2CO} concentrations from aqueous photolytic production peak at 1.5 nM, which is 3--4 orders of magnitude higher than the maximum \ce{H2CO} concentration from atmospheric rain-out (4.1$\times$10$^{-5}$ nM). On the other hand, for our late Hadean (oxidizing) model B, the \ce{H2CO} concentration from rain-out is 0.34 nM, which is 2 orders of magnitude higher than the \ce{H2CO} concentration from photolytic production. Therefore, in oxidizing conditions \ce{H2CO} in WLPs may also come directly from the atmosphere.

Our model solves another main limitation of meteorites as a source of prebiotic nucleobases, in that cytosine and thymine are not present in meteorites \citep{2011PNAS..10813995C,Reference46,Reference106}. During the early Hadean, guanine, cytosine, uracil and thymine concentrations peak at 8.2, 1.4, 0.7 and 0.5 nM, respectively. 2-aminooxazole and ribose concentrations peak at 0.045, and 0.015 nM, respectively. Meteorites would have provided brief ($\sim$a few weeks) enhancements in nucleobase concentrations up to three orders of magnitude above the base abundance from {\it in situ} production.

\begin{figure}[!hbtp]
\centering
\includegraphics[width=\linewidth]{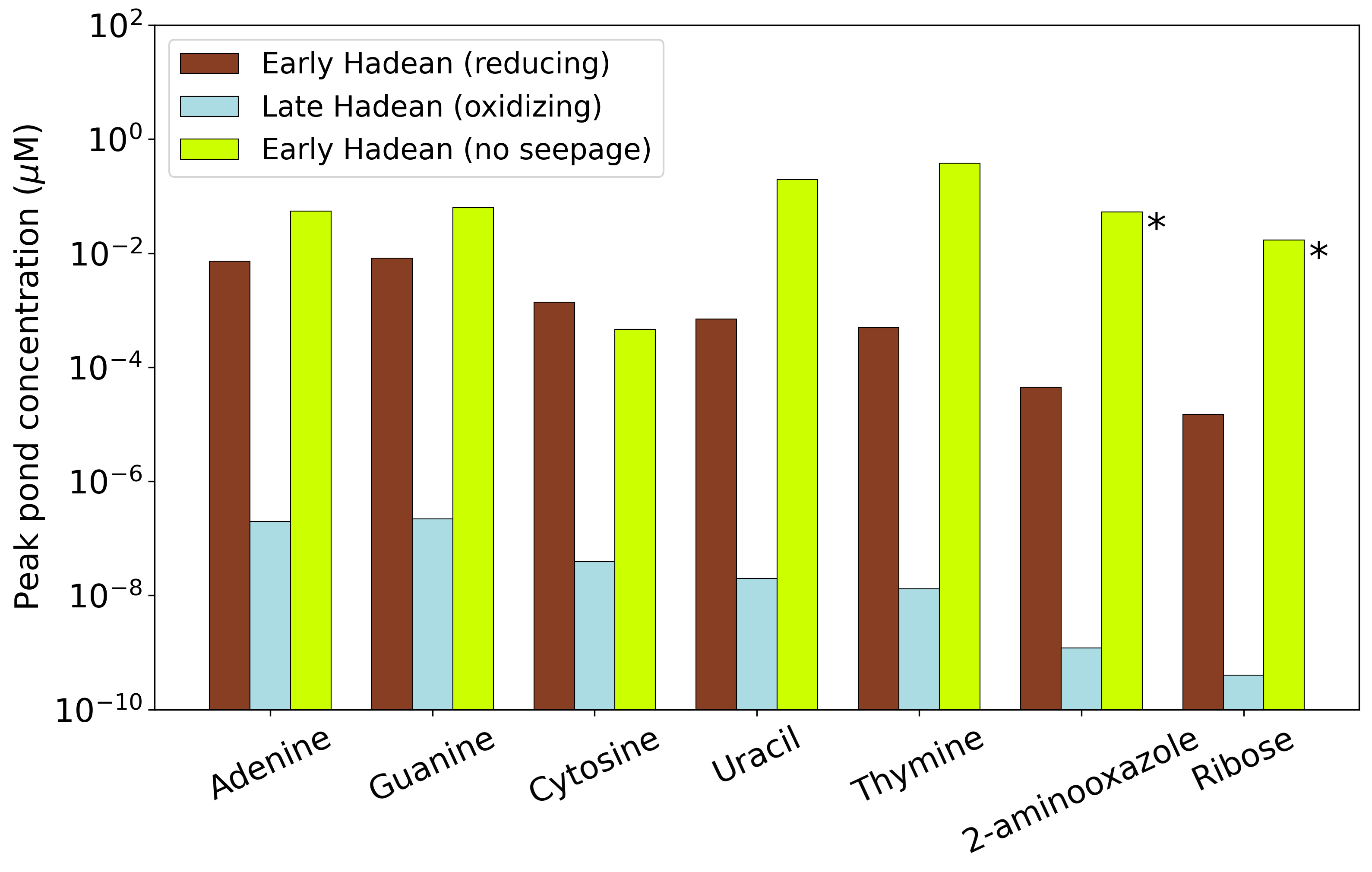}
\caption{Peak concentrations of biomolecules produced {\it in situ} in our WLP models (see Table~\ref{PeakConcentrations} for more details). *Hydrolysis rates are not estimated for 2-aminooxazole and ribose, therefore concentrations should be considered as maximum values in the absence of aqueous chemical sinks.}
\label{PeakConcFig}
\end{figure}

In Figure~\ref{PeakConcFig} and Table~\ref{PeakConcentrations}, we summarize the peak concentrations of key biomolecules and biomolecule precursors in our model WLPs for our early Hadean (reducing) and late Hadean (oxidizing) models A and B. We also include the early Hadean (reducing) model A with seepage turned off, representing a scenario where biomolecules are not exposed to this sink due to rock pore blockage by amphiphilic multilamellar matricies or mineral gels, or adsorption onto mineral surfaces. Concentrations of biomolecules from the early Hadean (reducing) model A are approximately 4 orders of magnitude higher than concentrations from the late Hadean (oxidizing) model B. When seepage is turned off, biomolecule concentrations increase in model A to $\sim$1--400 nM. It is interesting to note that there are order of magnitude differences in the concentrations of purine nucleobases (adenine and guanine), versus the pyrimidine nucleobases (cytosine, uracil and thymine). Does this have implications for the frequencies of various bases in the first replicating RNA polymers?

\section{Discussion}

\subsection{Comparison with other Methods and Experiments}

We have developed multiple new methods that have greatly enhanced the capabilities of non-equilibrium calculations of atmospheric HCN on early Earth. These include the calculation of composition-dependent pressure-temperature (P-T) profiles using a radiative transfer code, the inclusion of lightning chemistry and the time-dependent influx of \ce{H2}, \ce{CO2}, and \ce{CH4} from impact degassing, volcanism, and oceanic geochemistry, respectively.

Past non-equilibrium atmospheric models for the Archean ($\sim$3.8--2.5 bya) have computed HCN production for a range of \ce{CH4}, \ce{H2}, and \ce{CO2} abundances using a commonly suggested input P-T profile for early Earth with surface temperatures of 273--288 K \citep{Reference591,2011EPSL.308..417T}. These models imposed \ce{CH4} abundances in the 10--1000 ppm range, either did not fix \ce{H2} or imposed \ce{H2} abundances in the 0.01--1\% range, and imposed \ce{CO2} abundances in the 0.04--3\% range. Our strategy is different, as we begin with initial reducing (4.4 bya) or oxidizing (4.0 bya) conditions that are thought to represent each epoch, and calculate the composition-dependent input P-T profiles using a radiative transfer code. As a result, our surface temperatures are 27--78$^{\circ}$C hotter than these past Archean models. Then, we allow the concentrations of all species to evolve over time based on their source and sink rates at each epoch. 

For example, \ce{CO2} reaches a steady surface abundance of $\sim$0.05--0.1\% in our models based on the source rates from volcanic outgassing and association chemistry balanced with the sink rates from atmospheric rain-out and photodissociation. Our steady state abundances are similar to the lower end of \ce{CO2} abundances from past Archean models. Similarly, our end-of-simulation \ce{H2} abundances ( 18--19\% and ppb--ppm for reducing and oxidizing models, respectively) are balanced by the source rate from epoch-dependent impact bombardment, and the sink rates from hydrodynamic escape, photodissociation, and chemistry. The steady state \ce{H2} abundances in our reducing models are at least an order of magnitude higher than those calculated or imposed in the Archean models, while the steady state \ce{H2} abundances in our oxidizing models are at least two orders of magnitude lower than the values calculated or imposed in those models.

\ce{CH4} abundances are based on source rates from oceanic sources balanced with sink rates from photodissociation and chemistry. Surface \ce{CH4} abundances reach end-of-simulation values that are 0 to 4 orders of magnitude lower than the range of \ce{CH4} concentrations imposed in the past Archean models. Given the correlation between HCN production and \ce{CH4} abundance, our calculated surface HCN concentrations tend to be lower than those calculated in these Archean models. The Archean models compute HCN mixing ratios near the surface to be $\sim$10$^{-12}$--10$^{-7}$. The lowest end of this range is similar to the surface HCN abundances in our reducing 4.4 bya models of $\sim$10$^{-11}$, but three orders of magnitude higher than the HCN mixing ratios in our oxidizing 4.0 bya models of $\sim$10$^{-15}$ after $\sim$1 Myr. This three-order-of-magnitude discrepancy is due to the fact that our 4.0 bya models are more oxidizing than the previous Archean models. In other words, our 4.0 bya models contain higher concentrations of oxygen species such as \ce{O2} and \ce{H2O}, and lower concentrations of reducing gases such as \ce{H2} and \ce{CH4}.

\citet{Zahnle_et_al2020} modeled non-equilibrium chemistry from a post-large body impact with input P-T profiles based on a simple analytic equation for a moist adiabat. These models began with concentrations of \ce{H2}, \ce{CO}, \ce{CO2}, \ce{CH4}, and \ce{NH3} that result from various impactors equilibrating with different mineral redox buffers. They found that impactors at least Vesta in size (525 km) are required to sustain the high temperatures required for rapid methane production (e.g. 0.1--10 bars). These models produced post-impact atmospheric \ce{CH4} abundances of $\sim$3\%, which resulted in similarly high ($\sim$ a few \%) HCN abundances for a few million years after impact. These post-impact \ce{CH4} abundances are four orders of magnitude higher than the ppm-range steady state \ce{CH4} abundances from our reducing models.
The analytic equation \citet{Zahnle_et_al2020} used for obtaining a habitable surface and P-T profile for their non-equilibrium chemistry models does not consider the strength of various opacity sources such as \ce{H2}-\ce{H2} collisional induced absorption (CIA) and \ce{CH4}. \citet{Zahnle_et_al2020} compute surface temperatures of $\sim$320 K for a post-Vesta-impact atmosphere of 3.9 bars \ce{H2} and 0.17 bars of \ce{CH4}. We are unable to obtain habitable surfaces when modeling \ce{H2} atmospheres $>$2 bars using the equations of radiative transfer. In our radiative transfer models, we find \ce{H2}-\ce{H2} CIA produces a strong greenhouse effect above $\sim$1.13 bar of \ce{H2}. The surface temperature of our early Hadean (reducing) model A, which has 1.5 bars of \ce{H2}, 2 ppm \ce{CH4}, and ppm-range \ce{H2O}, is reaching the limits of habitability at 78$^{\circ}$C. This suggests that the resultant high atmospheric pressures of \ce{H2} and \ce{CH4} from the large-body impacts modeled by \citet{Zahnle_et_al2020} would produce atmospheric temperatures too hot for WLPs to exist. More research needs to be done to understand whether a post-large body impact could provide high HCN rain-out rates to WLPs after a substantial amount of \ce{H2} escapes from the upper atmosphere allowing the surface to cool to habitable temperatures.



Atmospheric models of nitrogen-rich rocky exoplanets that use C/O ratio as an adjustable parameter produce HCN mixing ratios of 10$^{-8}$--10$^{-7}$ for atmospheric C/O ratios near 0.5, and HCN mixing ratios of $\sim$10$^{-3}$ for C/O ratios $>$ 1.5 \citep{2019Icar..329..124R}. We do not use C/O as an adjustable parameter, as we find that the balance of outgassing and losses of species such as \ce{CO2}, \ce{H2O}, and \ce{CH4} in our models lead to surface C/O ratios that vary from $\sim$0.001--1 over the course of the simulations. These C/O ratios are generally lower than those explored by \citet{2019Icar..329..124R}.

\citet{Pinto_et_al1980} modeled the chemical kinetics of formaldehyde production in a primitive Earth atmosphere composed of \ce{N2}, \ce{H2O}, \ce{CO2}, \ce{CO}, and \ce{H2}. They employed a highly reduced network of 8 photochemical reactions and 31 neutral reactions between 13 species. Their computed formaldehyde rain-out rates ($\sim$10$^{11}$ mol yr$^{-1}$) are approximately 3 orders of magnitude higher than our maximum formaldehyde rain-out rates for Model B ($\sim$10$^8$ mol yr$^{-1}$). Part of this discrepancy is due to the higher efficiency of their \ce{H2CO} rain-out. They used a rain-out (scavenging) coefficient (cm$^3$s$^{-1}$) that is based on \ce{H2CO} rain-out on Earth today, whereas we compute rain-out using a deposition velocity (cm s$^{-1}$) that is calculated from a thin-film paramaterization model \citep{Wagner_et_al2002} and is used in more recent atmospheric models that follow the deposition velocity treatment \citep{Ranjan_2020,Hu_et_al2012}. The treatment used in \citet{Pinto_et_al1980} produces a \ce{H2CO} rain-out rate that is about a factor of 20 greater than our models would produce if we had the same surface \ce{H2CO} density. Considering these differences, we may be underestimating \ce{H2CO} rain-out by up to a factor of $\sim$20. However, the main source of discrepancy is because their atmosphere is more oxidizing than all of our models. They do not include \ce{CH4} or other reduced carbon species (e.g. \ce{C2H2}, \ce{C2H6}) in their network, which, if given initial abundances, would change the oxidation state of the atmosphere and decrease the production of oxidized carbon species such as \ce{H2CO}.


Miller-Urey experiments have shown that reducing conditions are more favorable than oxidizing conditions for biomolecule production \citep{Schlesinger_Miller1983,Cleaves_et_al2008}. For example, \citet{Schlesinger_Miller1983} found a $\sim$3--4 order of magnitude difference in amino acid yields when switching from reducing (\ce{H2}-dominant) to oxidizing (\ce{CO2}-dominant) experimental conditions. This is consistent with our results, where we have shown that the early reducing phase of the Hadean eon at 4.4 bya produces atmospheric HCN and RNA building blocks $\sim$4 orders of magnitude higher in concentration than during the late oxidizing phase at 4.0 bya. Some Miller-Urey experiments have shown reasonable success of amino acid production in \ce{CO2}/\ce{N2} conditions when buffering the solution with calcium carbonate \citep{Cleaves_et_al2008}; however, this is likely demonstrating catalytic effects that increase amino acid production yields from low HCN concentrations.



\subsection{Biomolecule Concentrations}

Of critical importance is how high pond concentrations of biomolecules such as nucleobases, ribose, and 2-aminooxazole would need to be in order for nucleotide synthesis and subsequent RNA polymerization to occur. Published laboratory experiments that react nucleobases to produce nucleosides and nucleotides typically use 400$\mu$M--100mM concentrations of nucleobases and obtain 0.01--74\% yields \citep{1963Natur.199..222P,Reference82,Saladino_et_al2017,Nam_et_al2018}. The lowest end of this experimental concentration range is three orders of magnitude higher than the maximum nucleobase concentrations from our early Hadean (4.4 bya) model without seepage ($\sim$400 nM). Mechanisms to increase nucleobase concentrations in WLPs would likely be necessary given that experiments at nanomolar concentrations to produce nucleotides and RNA have not been reported, and achieving high yields requires optimal conditions. Such mechanisms include adsorption to mineral surfaces, and sequestration into amphiphilic multilamellar matrices or mineral gels \citep{Damer_Deamer2019,Deamer2017}. 

The porosity of WLPs is difficult to assess. While something is known about the porosity of volcanic basalts \citep{Hamouda_et_al2014}, pond environments are likely to be messier affairs. Various kinds of debris and films would be expected to accumulate at the bottom of ponds, such as
amphiphilic multilamellar matrices or mineral gels. These would clog the pores, reducing losses due to seepage.  It is difficult to estimate how large this effect would be. We can however, estimate an upper limit to abundances of biomolecules by computing models for which seepage is set to zero. Turning off seepage in our models provides a 1--3 order of magnitude increase in nucleobase concentrations after 500--10,000 years (see Figure~\ref{NoSeepage}). In this scenario, hydrolysis takes over as the the main concentration-limiting sink.


Concentrations of ribose and 2-aminooxazole in our models are the most dilute of the RNA building blocks, in the sub-nM range. Again, laboratory experiments use much higher concentrations than this for nucleotide synthesis, typically in the mM to M range \citep{1963Natur.199..222P,Reference82,Nam_et_al2018,Reference56}. \citet{Saladino_et_al2017} used only 8$\mu$M ribose  to produce nucleosides in a neat formamide solution; however, because this experiment was performed with a different solvent, it is unclear whether similar results could occur in aqueous solution. Our results encourage additional experimental work to determine the levels that are sufficient for nucleotide synthesis in realistic prebiotic conditions.

\section{Conclusion}

Our comprehensive treatment of the early Earth's coupled atmosphere-impactor-ocean system reveals several striking insights. It is the initial high rate of \ce{H2} impact degassing soon after the moon forming impact with the Earth that keeps its atmosphere in a chemically reducing state. Photochemistry-dominated HCN formation from methane in the lower atmosphere rains out into WLPs steadily over about a hundred million years. There, aqueous chemistry continuously drives nucleobase and perhaps nucleotide precusor synthesis to levels that polymerization by condensation reactions can occur. Overall, this steady input totally dominates that which is possible from more isolated meteoritic infall events. With a declining bombardment throughout the Hadean eon \citep{Reference425,1990Natur.343..129C}, the transition from reducing to oxidizing atmospheric conditions is roughly linear from 4.4 to 4.0 bya. After 4.3 bya, the reducing conditions dissipate. This terminates new biomolecular formation so that RNA-based life would have already had to appear. The astrophysical and chemical processes we model are quite general. They are intrinsic to the late phases of terrestrial planet formation, anywhere. This suggests that life on Earth, and perhaps also on other Earth-like worlds, began in the chaotic conditions that prevailed soon after their formation.   





\section*{Acknowledgments}

We thank the anonymous referee whose comments and questions led to improvements in this work. We thank David Catling for his critical analysis of an early draft of this paper and Nick Wogan for catching an error in our pre-published calculations. We thank Steve Benner, Jeffrey Bada, David Deamer, Bruce Damer, Oliver Trapp, Paul Higgs, Greg Slater and Paul Rimmer for interesting discussions regarding this work. This research is part of a collaboration between the Origins Institute at McMaster University and the Heidelberg Initiative for the Origins of Life. The research of BKDP was supported by an NSERC Alexander Graham Bell Canada Graduate Scholarship-Doctoral (CGS-D) and Ontario Graduate Scholarship. KM was supported by the Excellence Cluster ORIGINS which is funded by the Deutsche Forschungsgemeinschaft (DFG, German Research Foundation) under Germany's Excellence Strategy - EXC-2094 - 390783311. REP is supported by an NSERC Discovery Grant. TH acknowledges support from the European Research Council under the Horizon 2020 Framework Program via the ERC Advanced Grant Origins 83 24 28. We acknowledge Compute Canada for allocating the computer time required for this research. We thank Steve Janzen at McMaster Media Resources for his superb work in rendering Figure 1.

\section*{Author Contributions}

B.K.D.P., K.M., R.E.P., and T.H. designed research; B.K.D.P. performed research; K.M. performed atmospheric chemistry simulations; B.K.D.P. performed numerical pond simulations; K.E.C. and B.K.D.P. performed P-T profile simulations; B.K.D.P. and R.E.P. analyzed data; and B.K.D.P., R.E.P., and T.H. wrote the paper.

\section*{Competing Interests}

No competing financial interests exist.

\beginappendixA

\section*{Appendix A - Supplementary Methods}

\subsection*{Atmospheric Simulations}

Self-consistent disequilibrium atmospheric simulations are carried out iteratively using the consistent reduced atmospheric hybrid chemical network oxygen extension (CRAHCN-O) \citep{Pearce2020b,Pearce2020a} coupled with a 1D chemical kinetic model (ChemKM) \citep{Molaverdikhani_2019,Molaverdikhani2020}, with input pressure-temperature (P-T) structures calculated using petitRADTRANS \citep{2019AA...627A..67M}.

CRAHCN-O now contains 259 one-, two-, and three-body reactions, whose rate coefficients are gathered from experiments when available ($\sim$40\%), and are otherwise calculated using accurate, consistent, theoretical quantum methods ($\sim$60\%). Approximately 93 of the reactions in CRAHCN-O were missing from the literature prior to their discovery in \citet{Reference598,Pearce2020b,Pearce2020a}. This network can be used to calculate HCN and \ce{H2CO} chemistry in atmospheres characterized by any of \ce{N2}, \ce{CO2}, \ce{CH4}, \ce{H2O}, and \ce{H2}. We added 28 new reactions to CRAHCN-O after our preliminary simulations showed the artificial build up of species that previously had no reaction sinks. We list the new two-body and three-body reactions in Tables~\ref{addedtwobodychemistry} and \ref{newthreebody}, respectively.

All values are experimental, except for the low- and high-pressure limit rate coefficients for \ce{C2H + H + M -> C2H2 + M} and the high-pressure limit rate coefficient for \ce{C + H2 + M -> ^3CH2 + M}, as there were no experimental values available. For these reactions, we calculate the rate coefficients using the same validated theoretical and computational quantum methods developed in \citet{Pearce2020b,Pearce2020a} for the other three-body reactions in CRAHCN-O.

\begin{longtable*}{lcccc}
\caption{New two-body reactions added to CRAHCN-O for our early Earth atmospheric models, and their experimental Arrhenius coefficients. These are the most efficient sink reactions for species that would otherwise erroneously build up over ten-to-hundred million year timescales. The Arrhenius expression is $k(T) = \alpha \left(\frac{T}{300}\right)^{\beta} e^{-\gamma/T}$. \label{addedtwobodychemistry}} \\
Reaction Equation & $\alpha$ & $\beta$ & $\gamma$ & Source(s) \\ \hline \\[-2mm]
 \ce{HCCO + NO -> HCNO + CO} & 1.4$\times$10$^{-11}$ & 0 & -320 & \citet{Carl_et_al2002}\\    
  \ce{HCCO + NO -> HCN + CO2} &  6.1$\times$10$^{-12}$ & -0.72 & -200 & \citet{Carl_et_al2002} \\
   \ce{HCCO + ^3O -> CO + CO + H} & 1.6$\times$10$^{-10}$  & 0 & 0 & \citet{Reference451} \\
    \ce{HCCO + H -> CO + 3CH2} & 2.1$\times$10$^{-10}$  & 0 &  0  & \citet{Glass_et_al2000}, \citet{Frank_et_al1988} \\
 \ce{NCO + O2 -> CO2 + NO} &  1.3$\times$10$^{-12}$ &   0 &  0  & \citet{Schacke_et_al1974} \\
 \ce{NCO + NO -> N2 + CO2} &  1.6$\times$10$^{-11}$  &   0  &   0   & \citet{Cooper_et_al1993}, \citet{Cooper_et_al1992} \\
  \ce{NCO + ^3O -> NO + CO} &   6.4$\times$10$^{-11}$ & -1.14 &   0  & \citet{Becker_et_al2000} \\
 \ce{NCO + H -> NH + CO} &  2.2$\times$10$^{-11}$  &  0   &   0  & \citet{Becker_et_al2000} \\
 \ce{HO2 + ^3O -> O2 + OH} &  5.4$\times$10$^{-11}$ & 0  & 0 & \citet{Reference451} \\
 \ce{HO2 + OH -> H2O + O2} &  4.8$\times$10$^{-11}$  & 0 &  -250 & \citet{Reference451} \\
  \ce{HO2 + H -> H2O + ^3O}  & 5.0$\times$10$^{-11}$  & 0  &  866 & \citet{Reference451} \\
 \ce{HO2 + H -> O2 + H2}  & 7.1$\times$10$^{-11}$ &  0 &  710 & \citet{Reference451} \\
  \ce{HO2 + H -> OH + OH} &  2.8$\times$10$^{-10}$ &  0 &  440 & \citet{Reference451} \\
\ce{O2 + HCO -> CO + HO2} &   8.5$\times$10$^{-11}$ &  0 &  850 & \citet{Reference509} \\
 \ce{O2 + C2H -> HCCO + ^3O} &  1.0$\times$10$^{-12}$ &  0 &  0 & \citet{Reference509} \\
 \ce{O2 + C2H -> HCO + CO} &  4.0$\times$10$^{-12}$ &  0 &  0 & \citet{Reference509} \\
 \ce{O2 + CN -> NCO + ^3O}  &  1.1$\times$10$^{-11}$ &  0 &  -205 & \citet{Reference451} \\
   \ce{O2 + ^4N -> NO + ^3O} &  4.5$\times$10$^{-12}$ &  1.0 &  3720 & \citet{Baulch_et_al1994} \\
  \ce{O2 + CH -> OH + CO} &  5.0$\times$10$^{-11}$ &  0 &  0 &  \citet{Lichtin_et_al1984,Lichtin_et_al1983} \\
 & & & & \citet{Duncanson_Guillory1983}, \citet{Messing_et_al1979} \\
 \ce{O2 + C -> CO + ^3O} &  3.0$\times$10$^{-11}$ &  0 &  0 & \citet{Geppert_et_al2000}, \citet{Dorthe_et_al1991}\\
 & & & & \citet{Becker_et_al1988}, \citet{Husain_Young1975} \\
 & & & & H\citet{Husain_Kirsch1971}, \citet{Braun_et_al1969} \\
 & & & & \citet{Martinotti_et_al1968} \\
 \ce{NO + ^4N -> N2 + ^3O} &   3.1$\times$10$^{-11}$ &  0 &  0 & \citet{Atkinson_et_al1989} \\
 \ce{NO + ^2N -> N2 + ^3O} &  6.0$\times$10$^{-11}$ & 0 & 0 & \citet{Reference549} \\
  \ce{NO + C -> CN + ^3O} &  2.5$\times$10$^{-11}$ &  0 & 0 & \citet{Baulch_et_al1994} \\
  \ce{C2H + CH4 -> C2H2 + CH3}  & 3.0$\times$10$^{-12}$ &  0 &  250 & \citet{Reference509} \\
 \ce{C2H + ^3O -> OH + CO} &  1.7$\times$10$^{-11}$ &  0 & 0 & \citet{Reference451} \\
\hline
\end{longtable*}

\begin{longtable*}{lccc}
\caption{New three-body reactions added to CRAHCN-O for our early Earth atmospheric models, and their calculated or experimental Lindemann coefficients. These are the most efficient high-pressure sink reactions for species that would otherwise erroneously build up over ten-to-hundred million year timescales. Experimental rate coefficients are listed when available, otherwise we calculate them using canonical variational transition state theory and Rice--Ramsperger--Kassel--Marcus/
master equation theory at the BHandHLYP/aug-cc-pVDZ level of theory (see \citet{Pearce2020b,Pearce2020a} for details on how these calculations are performed. $k_{\infty}$ is the second-order rate coefficient in the high pressure limit with units cm$^{3}$s$^{-1}$. $k_0$ is the third-order rate coefficient in the low pressure limit with units cm$^{6}$s$^{-1}$. These values fit into the pressure-dependent rate coefficient equation $k$ = $\frac{k_0 [M] / k_{\infty}}{1 + k_0 [M] / k_{\infty}} k_{\infty}$. \label{newthreebody}} \\
Reaction equation & k$_\infty$(298) &  k$_0$(298) & Source(s) \\ \hline \\[-2mm]
  \ce{O2 + H + M -> HO2 + M} & 7.5$\times$10$^{-11}$ & (M=\ce{N2}) 3.9$\times$10$^{-30}$ T$^{-0.8}$ & (\ce{N2}) \citet{Reference451} \\ 
  & & (\ce{CO2}) 5.8$\times$10$^{-30}$ T$^{-0.8}$ & (\ce{H2O}) \citet{Reference451} \\
    & & (\ce{H2}) 4.3$\times$10$^{-30}$ T$^{-0.8}$ & (\ce{H2}) \citet{Reference451}\\
 \ce{C2H + H + M -> C2H2 + M} & 2.3$\times$10$^{-11}$ & (M=\ce{N2}) 5.8$\times$10$^{-28}$ & This work \\   
 & & (\ce{CO2}) 7.1$\times$10$^{-28}$ & \\
  & & (\ce{H2}) 4.2$\times$10$^{-28}$ & \\
  \ce{C + H2 + M -> ^3CH2 + M} & 1.6$\times$10$^{-9}$ & (M=\ce{N2}) 7.0$\times$10$^{-32}$ & This work, \citet{Husain_Young1975}, \\ 
  & & (\ce{CO2}) 7.0$\times$10$^{-32}$ & \citet{Husain_Kirsch1971} \\
    & & (\ce{H2}) 7.0$\times$10$^{-32}$ & \\
 \hline
\end{longtable*}

The ChemKM code takes as input: A) a chemical network, B) an atmospheric pressure-temperature (P-T) structure, C) an eddy diffusion profile to characterize turbulent mixing, D) the solar radiation spectrum at the top-of-atmosphere (TOA), E) wavelength-dependent photochemical reactions, and F) incoming and outgoing molecular fluxes from the surface and TOA (i.e. impact degassing, volcanic outgassing, rain-out, hydrogen escape, and chemical production from lightning). ChemKM uses the plane-parallel two-stream approximation to calculate radiative transfer, and includes both photoabsorption and Rayleigh scattering. The pressure profiles remain static throughout the simulations; therefore we must assume that pressure has reached equilibrium with the influx and outflux of atmospheric gases. This assumption is valid for our models, as our \ce{H2} impact degassing rates never exceed the \ce{H2} escape rates.

P-T structures for Early Earth models are calculated using the petitRADTRANS software package \citep{2019AA...627A..67M}. petitRADTRANS is a 1D radiative transfer code that uses the Guillot analytic temperature model \citep{2010AA...520A..27G} and correlated-k opacity tables to solve for atmospheric temperatures of planets with no surface boundary condition. Visible opacities are calculated using the Planck mean, and infrared opacities are calculated using the Rosseland mean \citep{2014AA...562A.133P}. Our models are all 100 layers, from surface pressures of $\sim$1--2 bar to TOA pressures of 1$\times$10$^{-8}$ bar. We implement cloudless and hazeless models given the large uncertainties of these parameters for the early atmosphere. We note that the lack of biogenic CCN would have led to a shorter lifetime for optically thick convective clouds during the Hadean \citep{Rosing2010}, reducing their contribution to both the greenhouse and albedo when compared with those of modern Earth \citep{Charnay_et_al2020}. 

\begin{figure*}[!hbtp]
\centering
\includegraphics[width=\linewidth]{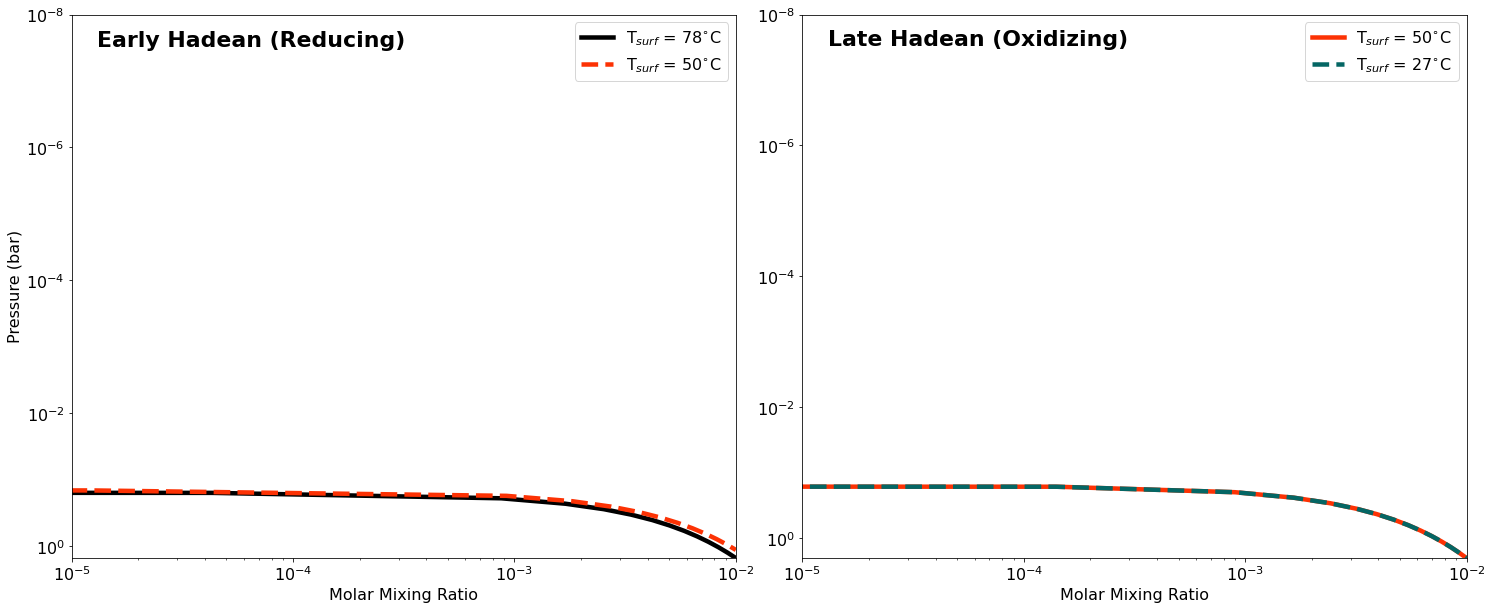}
\caption{Initial tropospheric water vapor profiles for our 4 early Earth models.}
\label{water_profiles}
\end{figure*}

In Figure~\ref{water_profiles} we display the initial water vapor profiles in the troposphere of our four early Earth models. We make an incremental improvement over the standard \citet{Manabe_Wetherald1967} water vapor profile for Earth's atmosphere by calculating tropospheric water abundances in two steps. In the first step, we iterate the Arden Buck equation \citep{Buck1981}, which is dependent on temperature, and the P-T calculations from petitRADTRANS, which are dependent on water composition. To avoid a runaway greenhouse due to water vapor feedback, we parameterize the strength of water vapor feedback by decreasing the relative humidity (RH) by $\beta$ = 6\% for every $^{\circ}$C of warming \citep{Held_Soden2000}. In step two, we smooth out these profiles to avoid numerical instabilities in ChemKM. We use the calculated water vapor at the tropopause and linearly increase the mixing ratio moving downwards in altitude to reach a typical water vapor abundance of 1\% for wet rocky planets \citep{Hu_et_al2012}.
The tropopause is chosen to be 0.14 bar, similar to the present day Earth \citep{Robinson_Catling2014}.

The three main inputs for our P-T structure calculations are A) input composition, B) equilibrium temperature (T$_{\mathrm{eq}}$), and C) internal temperature (T$_{\mathrm{int}}$). Initial guesses for input compositions were selected to represent reducing (\ce{H2}/\ce{N2}-dominant) or oxidizing (\ce{CO2}/\ce{N2}-dominant) phases of the early (4.4 bya) and late (4.0 bya) Hadean eon. Surface pressure and methane abundance are adjusted from 1--2 bar and 1--10 ppm, respectively, to maintain habitable surface temperatures (i.e. 0 $^{\circ}$C $\leq$ T$_{\mathrm{s}}$ $\leq$ 100 $^{\circ}$C) (see Table~\ref{models}). Equilibrium temperatures are calculated using the equation below.

\begin{equation}
T_{\mathrm{eq}} = \left( \frac{(1-A) L_{\odot}}{16 \pi \sigma a^2} \right)^{1/4},
\end{equation}
where T$_{\mathrm{eq}}$ is equilibrium temperature, $A$ is albedo, $L_{\odot}$ is solar luminosity, $\sigma$ is the Stefan-Boltzmann constant, and $a$ is the semi-major axis of the planet.

Luminosities for the Sun at 4.4 bya (0.705$L_{\odot}$) and 4.0 bya (0.728$L_{\odot}$) are obtained from a pre-computed stellar evolution model of a Sun-like star \citep{Heller2020,2015AA...577A..42B}. The Hadean Earth would have been mostly covered in water; therefore, albedo is taken to be 0.06, which is consistent with a cloudless water world \citep{Roesch2002}.

Internal heat flow is taken to be three times the present value, which is compatible with thermal modeling of the Hadean \citep{Reference89}. Using the Stefan-Boltzmann law, this results in internal temperatures of T$_{\mathrm{int}}$ = 43.3 K.

\begin{figure*}[!hbtp]
\centering
\includegraphics[width=\linewidth]{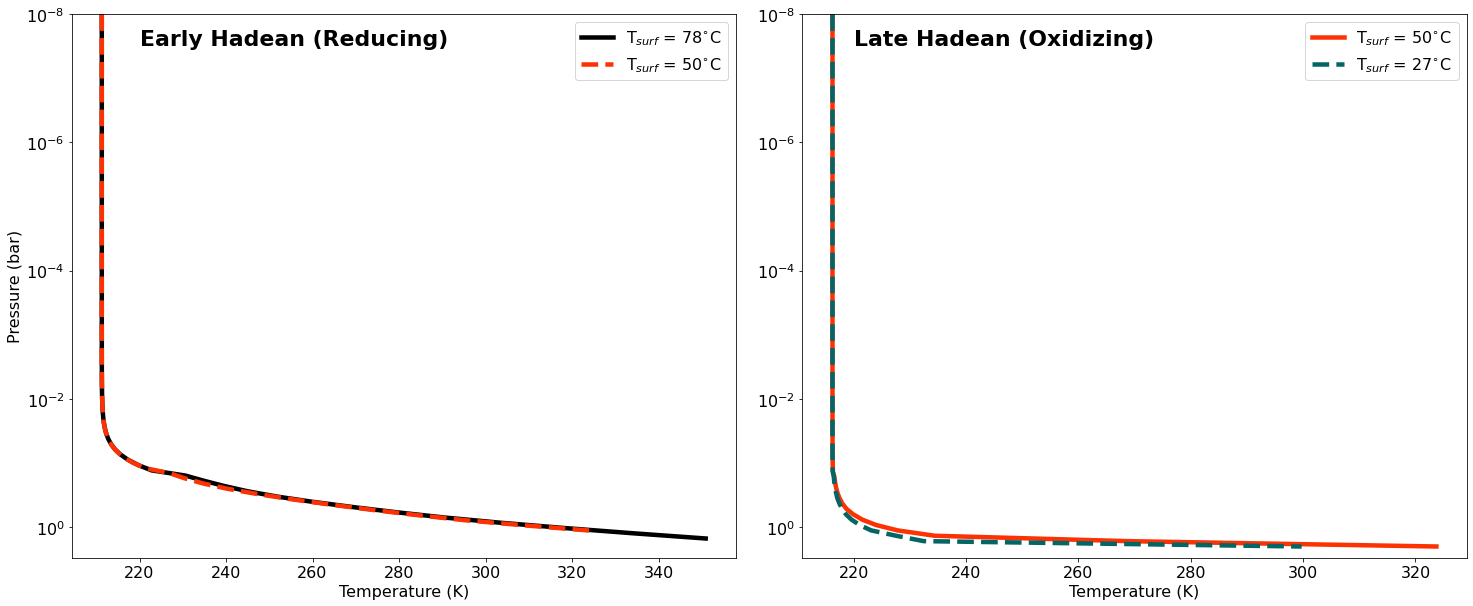}
\caption{Pressure-Temperature profiles for our 4 early Earth models, calculated with petitRADTRANS using the input compositions displayed in Table~\ref{models}. }
\label{P-T}
\end{figure*}

In Figure~\ref{P-T}, we display the P-T profiles for our four Hadean models.

\begin{figure}[!hbtp]
\centering
\includegraphics[width=\linewidth]{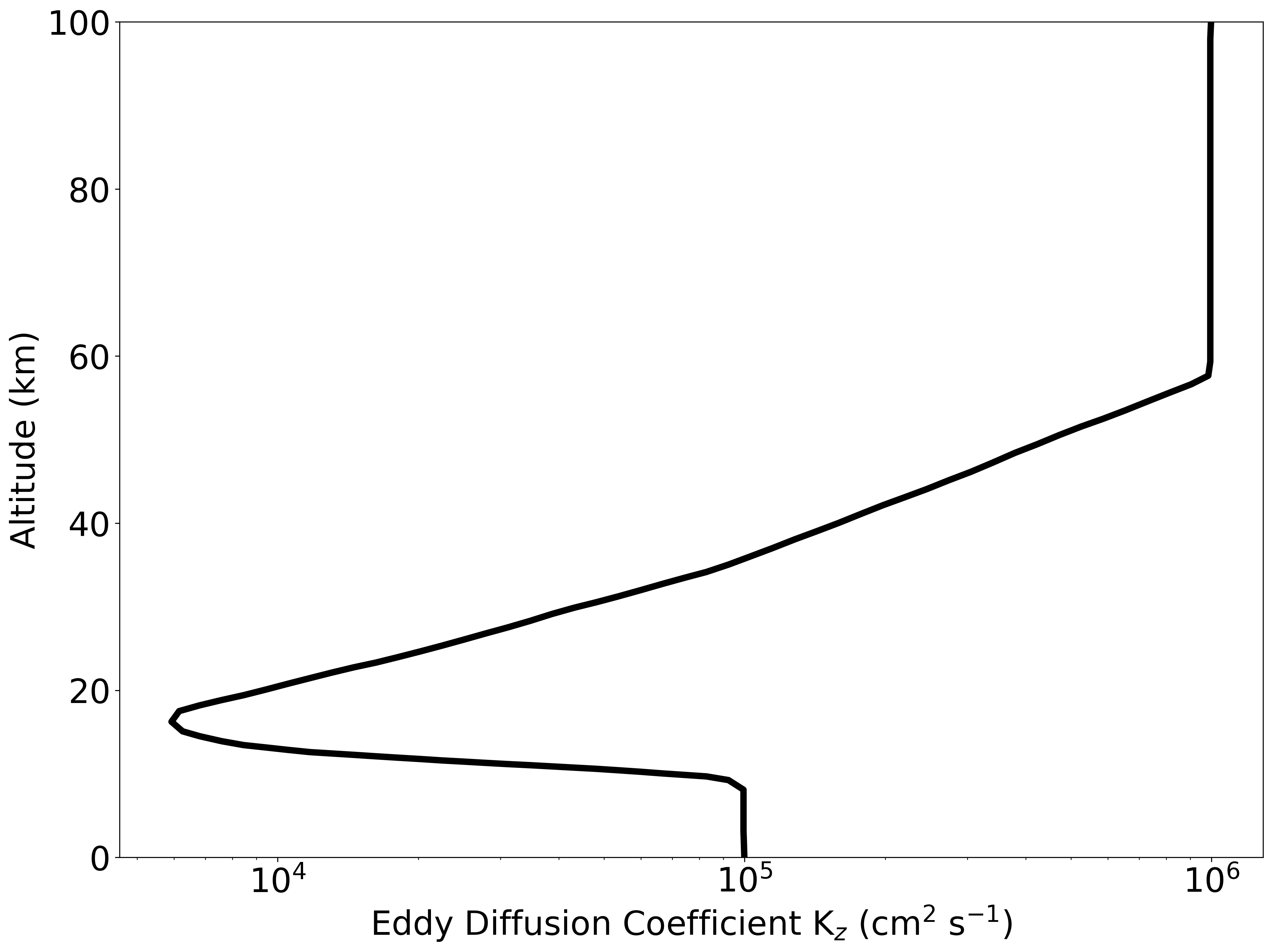}
\caption{Eddy diffusion profile for all early Earth models, characterizing turbulent mixing in the atmosphere. This is the standard K$_{\mathrm{z}}$ profile used for atmospheric simulations of early Earth and analogous exoplanets \citep{Ranjan_2020,Reference124,2011EPSL.308..417T,Reference591}.}
\label{eddy}
\end{figure}

In Figure~\ref{eddy}, we display the eddy diffusion profile used for all our early Earth modes. This is the standard profile for early Earth and analogous exoplanets \citep{Ranjan_2020,Reference124,2011EPSL.308..417T,Reference591}.

The TOA radiation for our models is based on the solar mean \citep{2004AdSpR..34..256T} with a solar zenith angle of 50$^{\circ}$; however, we increase the UV flux to simulate the increased activity of the young Sun. In Table~\ref{UVirradiance}, we display the multiplicative factors used in our models for each UV wavelength interval. Values are based on observations of young solar analogs (ages $\sim$0.1--7 Gyr) \citep{Ribas_et_al2005}.

\begin{table}[ht!]
\centering
\caption{Multiplicative increase in UV irradiation from present day values based on observations of young solar analogs \citep{Ribas_et_al2005}. \label{UVirradiance}} 
\begin{tabular}{lcccc}
\\
\multicolumn{1}{l}{Model} &
\multicolumn{1}{c}{1--20 $\AA$} &
\multicolumn{1}{c}{20--360 $\AA$} &
\multicolumn{1}{c}{360--920 $\AA$} &
\multicolumn{1}{l}{920--1200 $\AA$} 
\\[+2mm] \hline \\[-2mm]
4.0 bya & 60 & 10 & 9 & 7  \\
4.4 bya & 500 & 60 & 30 & 20 \\
\hline
\end{tabular}
\end{table}

Our 33 photochemical reactions mostly match those from the Titan models by \citet{2012AA...541A..21H}; however, we update the \ce{H2O} absorption cross-sections with the recent near-UV experimental measurements from \citet{Ranjan_2020}, we remove erroneous \ce{CO2} absorption below 202 nm \citep{Ranjan_2020}, and we add photochemistry for \ce{O2} and \ce{HO2} following treatments in \citet{Hu_et_al2012} and \ce{CH3OH} using experimental cross-sections from \citet{Lange_et_al2020} and \citet{Burton_et_al1992}.

We couple planetary surface processes to our atmospheric models by adding influxes of species to the lowest layer of our atmospheres. These include: \ce{H2} impact degassing, \ce{CO2} outgassing from volcanoes, \ce{CH4} outgassing from hydrothermal systems, \ce{H2O} evaporation from the ocean, and chemical production (e.g. HCN, CO, \ce{^3O}, H) due to lightning.

\begin{figure}[!hbtp]
\centering
\includegraphics[width=\linewidth]{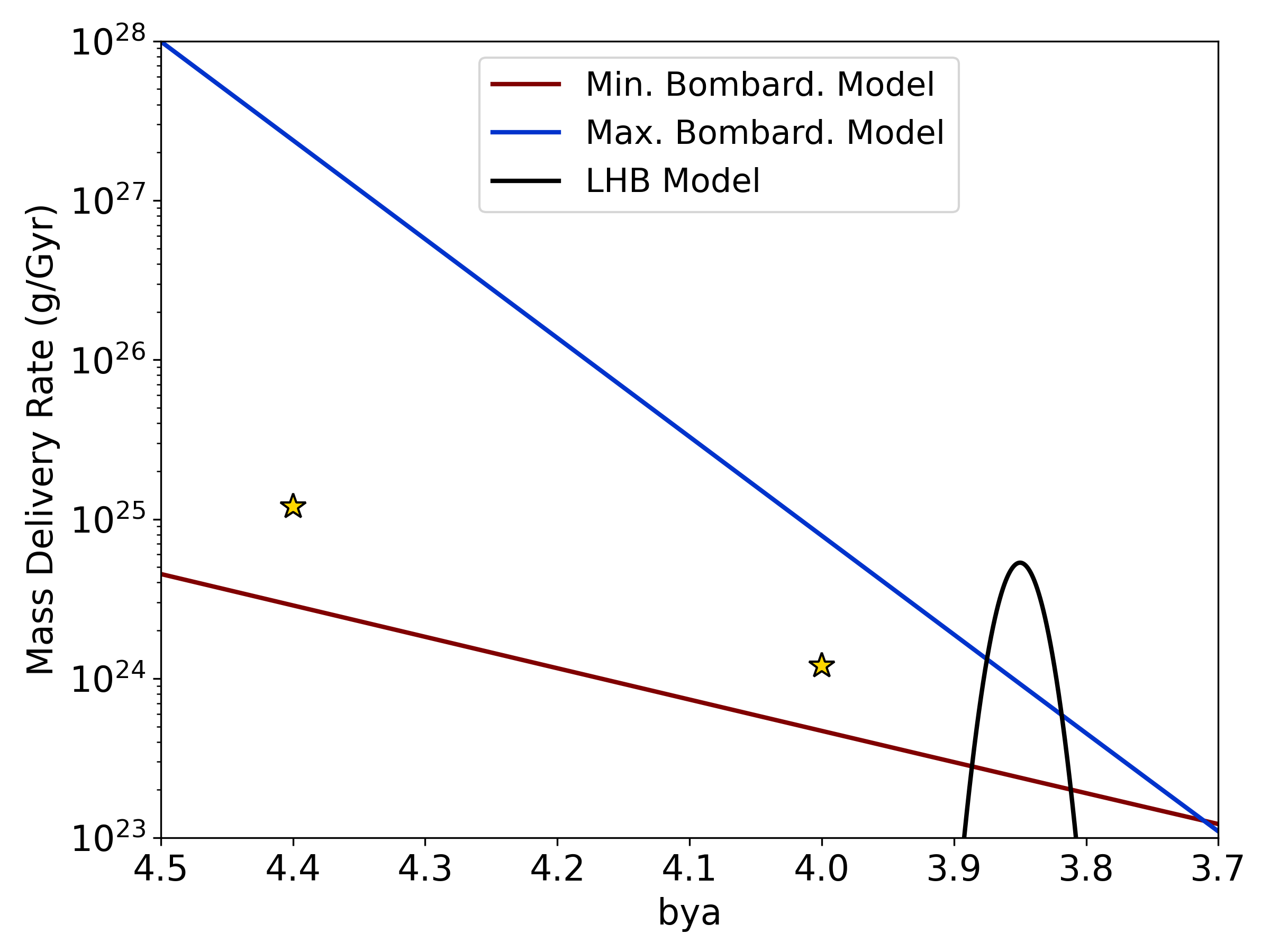}
\caption{Three early Earth bombardment models from \citet{Reference425} based on fits to the lunar cratering record. The gold stars represent the bombardment rates we used to calculate \ce{H2} impact degassing for our 4.4 and 4.0 bya atmospheric models.}
\label{bombard}
\end{figure}

Equilibrium chemistry calculations performed by \citet{Zahnle_et_al2020} for enstatite chondrite impactors suggest that \ce{H2} degassing via the reaction \ce{Fe + H2O -> FeO + H2} scales linearly with impactor mass, at a rate of $\sim$ 10$^{-21}$ mol \ce{H2} cm$^{-2}$ g$^{-1}$ impactor. Mathematical fits to the lunar cratering record provide us with an estimate of the rate of impactors on early Earth at a given epoch. In Figure~\ref{bombard}, we display our model bombardments rates, which lie between the minimum and maximum bombardment fits \citep{Reference425,1990Natur.343..129C}. These bombardment rates are 1.2$\times$10$^{25}$ g Gyr$^{-1}$ and 1.2$\times$10$^{24}$ g Gyr$^{-1}$ at 4.4 bya and 4.0 bya, respectively. Multiplying the \ce{H2} degassing abundance per unit mass by the mass delivery rates at 4.4 bya and 4.0 bya gives us \ce{H2} impact degassing rates of 2.3$\times$10$^{11}$ cm$^{-2}$ s$^{-1}$ and 2.3$\times$10$^{10}$ cm$^{-2}$ s$^{-1}$ for the early and late Hadean models, respectively.

We model H and \ce{H2} escape using the approximation developed by \citet{Zahnle_et_al2019,Zahnle_et_al2020} which blends energy-limited and diffusion-limited escape. The equation is

\begin{equation}
\left(\frac{dN_{H_2}}{dt}\right)_{esc} = -\frac{A S}{\sqrt{1 + B^2 S^2}} \frac{N_{H_2}}{\Sigma_j N_j}   (cm^{-2} s^{-1})
\end{equation}
where A = 2$\times$10$^{12}$ cm$^{-2}$ s$^{-1}$, B$^2$ = 0.006, S is the XUV and FUV irradiation relative to modern Sun (i.e. 30 and 9 for 4.4 and 4.0 bya, respectively), N$_{H_2}$ is the number of \ce{H2} molecules, and N$_{j}$ is the number of molecules of species $j$. 

The limits of \ce{H2} degassing to assume a static pressure profile in equilibrium with \ce{H2} escape would be $\sim$2.3$\times$10$^{13}$ and 1.4$\times$10$^{13}$ cm$^{-2}$ s$^{-1}$ at 4.4 and 4.0 bya, respectively.

We use a constant \ce{CO2} outgassing rate of 3.0$\times$10$^{11}$ cm$^{-2}$ s$^{-1}$ in all our models, which is consistent with Earth-like volcanic outgassing used in other atmospheric models \citep{Hu_et_al2012}. 

\citet{Guzman-Marmolejo2013} modeled the production of \ce{CH4} in hydrothermal systems for an Earth-like planet, and calculated a production rate of 6.8$\times$10$^{8}$ cm$^{-2}$ s$^{-1}$. \citet{Guzman-Marmolejo2013} also modeled \ce{H2} production in hydrothermal systems; however, the rates of \ce{H2} produced in hydrothermal environments are orders of magnitude lower than the rates of \ce{H2} production from impact degassing.

We do not include loss of \ce{CH4} due to haze production, as our \ce{CH4}/\ce{CO2} ratio never exceeds 0.1 (which is a common identifier for haze production) \citep{Trainer_et_al2006}.

In Table~\ref{rain-out}, we list the deposition velocities for the species that are rained out of the lowest layer in our atmospheric models.

\begin{table}[ht!]
\centering
\caption{Deposition velocities for the chemical species in our early Earth models. \label{rain-out}} 
\begin{tabular}{lll}
\\
\multicolumn{1}{l}{Species} &
\multicolumn{1}{l}{Deposition (cm s$^{-1}$)} &
\multicolumn{1}{l}{Source} 
\\[+2mm] \hline \\[-2mm]
\ce{CO2} & 1$\times$10$^{-4}$ & \citet{Archer2010} \\
\ce{CH3OH} & 0.1 & \citet{Wohlfahrt_et_al2015} \\
\ce{O2} & 1$\times$10$^{-8}$ & this work \\
\ce{CO} & 1$\times$10$^{-8}$ & \citet{Kharecha_et_al2005} \\
\ce{H2O2} & 0.5 & \citet{Hauglustaine_et_al1994} \\
\ce{C2H6} & 1$\times$10$^{-5}$ & \citet{Hu_et_al2012} \\
\ce{HO2} & 1 & \citet{Ranjan_2020} \\
\ce{H2CO} & 0.1 & \citet{Wagner_et_al2002} \\
\ce{HCO} & 0.1 & \citet{Ranjan_2020} \\
\ce{HCN} & 7$\times$10$^{-3}$ & \citet{2011EPSL.308..417T} \\
\ce{OH} & 1 & \citet{Ranjan_2020} \\
\ce{^3O} & 1 & \citet{Ranjan_2020} \\
\ce{H} & 1 & \citet{Ranjan_2020} \\
 \hline
\end{tabular}
\end{table}

We apply a \ce{CO2} deposition velocity of 1.0$\times$10$^{-4}$ cm s$^{-1}$, which is estimated in \citet{Hu_et_al2012} to produce a \ce{CO2} lifetime consistent with the lifetime of silicate weathering on Earth. We use a CO deposition velocity of 1$\times$10$^{-8}$ cm s$^{-1}$, calculated from a 2-box model \citep{Kharecha_et_al2005}. We also use a deposition velocity of 1$\times$10$^{-8}$ cm s$^{-1}$ for \ce{O2}, given its similar solubility and diffusivity to CO \citep{Harman_et_al2015}. We use the standard HCN deposition velocity of 7.0$\times$10$^{-3}$ cm s$^{-1}$ that is used in other early Earth models \citep{2011EPSL.308..417T,Reference591}. Additional deposition velocities are chosen to be consistent with other rocky exoplanet atmospheric models \citep{Hu_et_al2012,Ranjan_2020}.
Major species not listed in this table (e.g. \ce{H2} and \ce{CH4}) are not very soluble in water, therefore we do not include rain-out for these species \citep{NISTChemWebBook}.

\subsubsection*{Lightning}\label{lightning}

Lightning chemistry in the context of the origin of life was first developed experimentally in the 1950's \citep{Miller1953}. The fundamental Miller-Urey experiment involves sending an electric discharge through a combination of reduced gases to trigger dissociation. The radicals produced in this process then react to form biomolecule precursors such as HCN and \ce{H2CO} \citep{Reference600,Bada2016}. These precursors condense into a reservoir, where aqueous chemistry produces biomolecules such as amino acids \citep{Miller1953,Reference440} and nucleobases \citep{Reference438}.

Present-day Earth has an average global lightning flash density of $\sim$2 flashes km$^{-2}$ yr$^{-1}$ \citep{Hodosan2016}. However, above just the oceans, this average density drops to 0.3--0.6 flashes km$^{-2}$ yr$^{-1}$. Given the smaller coverage of continental crust above sea water during the Hadean, we set the global lightning flash density for our models to 1 flash km$^{-2}$ yr$^{-1}$; however, we also explore an average lightning flash density measured locally during volcanic eruptions on Earth today ($\sim$10$^{4}$ flashes km$^{-2}$ yr$^{-1}$).

\begin{table}[ht!]
\centering
\caption{Equilibrium abundances (molar mixing ratios) from lightning chemistry occurring in our four early Earth models. Thermodynamic simulations are based on initial concentrations in Table~\ref{models} for a freeze out temperature of T$_F$ = 2000 K. \label{equilib}} 
\begin{tabular}{lcccc}
\\
\multicolumn{1}{l}{Species} &
\multicolumn{1}{c}{Model A} &
\multicolumn{1}{c}{Model B} &
\multicolumn{1}{c}{Model C} &
\multicolumn{1}{l}{Model D}
\\[+2mm] \hline \\[-2mm]
 \ce{HCN} & 1.4$\times$10$^{-10}$ & 1.8$\times$10$^{-13}$ & 5.1$\times$10$^{-11}$ & 1.8$\times$10$^{-13}$ \\
 \ce{H2} & 8.9$\times$10$^{-1}$ & 8.7$\times$10$^{-1}$ & 8.9$\times$10$^{-1}$ & 8.7$\times$10$^{-1}$ \\
 \ce{H}  & 1.3$\times$10$^{-3}$ & 6.0$\times$10$^{-6}$ & 1.4$\times$10$^{-3}$ & 6.0$\times$10$^{-6}$ \\
 \ce{N2} & 9.9$\times$10$^{-2}$ & 9.8$\times$10$^{-2}$ & 9.9$\times$10$^{-2}$ & 9.8$\times$10$^{-2}$ \\
\ce{^4N} & 2.3$\times$10$^{-10}$ & 2.0$\times$10$^{-10}$ & 2.7$\times$10$^{-10}$ & 2.0$\times$10$^{-10}$ \\
\ce{H2O} & 9.9$\times$10$^{-3}$ & 9.7$\times$10$^{-3}$ & 9.9$\times$10$^{-3}$ & 9.7$\times$10$^{-3}$ \\
\ce{^3O} & 1.4$\times$10$^{-9}$ & 3.4$\times$10$^{-5}$ & 1.9$\times$10$^{-9}$ & 3.4$\times$10$^{-5}$ \\
 \ce{OH} & 1.4$\times$10$^{-6}$ & 2.2$\times$10$^{-4}$ & 1.6$\times$10$^{-6}$ & 2.2$\times$10$^{-4}$ \\
 \ce{NO} & 1.6$\times$10$^{-8}$ & 4.6$\times$10$^{-4}$ & 1.9$\times$10$^{-8}$ & 4.6$\times$10$^{-4}$ \\
\ce{O2} & 6.9$\times$10$^{-12}$ & 5.3$\times$10$^{-3}$ & 9.1$\times$10$^{-12}$ & 5.3$\times$10$^{-3}$ \\
 \ce{CH4} & 1.4$\times$10$^{-11}$ & 4.2$\times$10$^{21}$ & 4.0$\times$10$^{-12}$ & 4.2$\times$10$^{-21}$ \\
 \ce{CH3} & 5.0$\times$10$^{-13}$ & 2.3$\times$10$^{-20}$ & 1.6$\times$10$^{-13}$ & 2.0$\times$10$^{-20}$ \\
 \ce{CO} & 2.0$\times$10$^{-6}$ & 1.1$\times$10$^{-2}$ & 9.9$\times$10$^{-7}$ & 1.1$\times$10$^{-2}$ \\
\ce{CO2} & 4.8$\times$10$^{-9}$ & 8.7$\times$10$^{-1}$ & 2.4$\times$10$^{-9}$ & 8.7$\times$10$^{-1}$ \\
\hline
\end{tabular}
\end{table}

We considered both non-equilibrium and equilibrium approaches for modeling lightning chemistry. For our non-equilibrium approach, we integrated the production of key radicals for the first $\sim$40$\mu$s of a lightning strike using the pressure and temperature evolution from \citet{Ardaseva2017}. However, this approach had accuracy issues as a complete high-temperature reaction network is required to accurately calculate the chemical evolution within a cooling lightning channel. This approach also led to some insensible results such as HCN production that is independent of lightning flash density.

Therefore, we use an equilibrium approach for modeling the lightning production of HCN and other species based on the lightning chemistry models for HCN and NO production by \citet{Chameides_Walker1981}. Lightning channels heat up to a point ($\sim$30000 K) \citep{Orville1968b} where the equilibrium timescale is less than 1$\mu$s \citep{Hill_et_al1980}. In fact, ab initio molecular dynamics simulations of electric discharges suggest the timescale for lightning chemistry may be on the order of picoseconds \citep{Cassone_et_al2018}. This is fast compared to the hundred millisecond cooling timescale of a lightning channel, as well as the 10$\mu$s eddy diffusion timescale in the lowest layer in our atmospheres.

Chemical abundances rapidly reach equilibrium while the lightning channel is above the freeze out temperature (T$_F$). The freeze out temperature is the temperature at which the concentration of a species can still be described by its equilibrium value. Beyond this point, there is not enough time at a given temperature for equilibrium to be reached, and thus the concentrations are frozen into the gas for the remainder of the cooling of the lightning channel. Reaction rate coefficients that break down a species are used to roughly determine the freeze out temperature, e.g., HCN + M → CN + H + M. Typical freeze out temperatures range from 1000–5000 K. The freeze out temperature for HCN is $\sim$2000--2500 K for lightning strikes similar in energy to Earth today (10$^5$ J m$^{-1}$) \citep{Chameides_Walker1981}. Other species such as NO have higher freeze out temperatures near 3000--3500 K for similar lightning discharge energies, but can also be $\sim$2000 K for the highest discharge energies (10$^{15}$ J m$^{-1}$). We adopt T$_F$ = 2000 K for our equilibrium calculations to estimate the mixing ratios for HCN and 13 other dominant equilibrium products in the early Earth lightning models by \citet{Chameides_Walker1981}.

Equilibrium calculations are performed using the thermochemical data from the JANAF tables \citep{Stull_Prophet1971}, and the ChemApp Software library (distributed by GTT Technologies, http://gtt.mch.rwth-aachen.de/gtt-web/).

In Table~\ref{equilib}, we display the equilibrium mixing ratios based on the initial abundances in our four early Earth models for a freeze out temperature of T$_F$ = 2000 K.

For our fiducial models, we introduce HCN and other species to the bottom layer of the atmosphere at a rate that corresponds to 1 flash km$^{-2}$ yr$^{-1}$. The species influx rates from lightning chemistry are calculated along a lightning channel extending through the first layer of the atmosphere using the following equation:
\begin{equation}
\frac{d[M]}{dA dt} = \frac{n_M P \Delta H \dot{f} \sigma_l}{k_B T \gamma} \left(\frac{1}{100}\right)^3 \left(\frac{1}{10000}\right)^2,
\end{equation}

where $\frac{d[M]}{dA dt}$ is the molar concentration of species M produced per cm$^{2}$ per second, $n_M$ is the molar mixing ratio of species M produced in the lightning strike (cm$^{-3}$), $\Delta H$ is the height of the lowest atmospheric layer (cm), $\dot{f}$ is the lightning flash density in flashes km$^{-2}$ yr$^{-1}$, $\sigma_l$ is the cross section of the lightning channel ($\sim$1 cm$^{2}$), $\gamma$ = 3,600 $\cdot$ 24 $\cdot$ 365.25 s yr$^{-1}$, and the remainder is unit conversion.

\subsection*{Warm little pond models}\label{methods2}

In Table~\ref{biomolecule_yields}, we display the sources and sinks for  nucleobases, ribose and 2-aminooxazole in our warm little pond models. All biomolecule reaction yields are based on HCN. For cytosine, uracil, thymine, ribose, and 2-aminooxazole, which require formaldehyde as a reactant, we use a formaldehyde yield from HCN of 3.6\%, which is three times the glyceronitrile yield from radiolytic aqueous HCN experiments performed by \citet{Yi_et_al2020}. We assume, to first-order, that the UV radiation incident at the pond surface allowed this aqueous photolytic reaction to proceed in similar yields as the \citet{Yi_et_al2020} radiolytic experiments.

\setlength\LTcapwidth{\textwidth}
\begin{longtable*}{lccccc}
\caption{Sources and sinks for the five nucleobases, ribose, and 2-aminooxazole in our warm little pond model. HCN enters our ponds from rain-out calculated in our antecedent atmospheric model, and is multiplied by experimental and theoretical yields to simulate the \emph{in situ} production of key RNA biomolecules. \ce{H2CO}, which is a key reactant for cytosine, uracil, and ribose synthesis is produced in our ponds directly from HCN \citep{Yi_et_al2020}. HCN reactions are fast (experiments last $<$ days) in comparison to the duration of our models. \label{biomolecule_yields}} \\
Biomolecule & Yield from HCN & Yield reference & Sinks & Sink rate & Sink reference\\ \hline \\[-2mm]
Adenine & 0.005--0.18$^a$ & \citet{Reference27}, & Photodestruction & 1.0$\times$10$^{-4}$ photon$^{-1}$ & \citet{2015AsBio..15..221P} \\
 & & \citet{2002OLEB...32...99H}, & & & \\
 & & \citet{ReferenceWaka} & & & \\
 & & &  Seepage & 2.6 mm solution d$^{-1}$ & \citet{Reference116}, \\
 & & & & & \citet{Reference425} \\
 & & & Hydrolysis & 5.0$\times$10$^{-10}$ s$^{-1}$ & \citet{Reference45}
 \\
  & & & & & \\
 Guanine & 6.7$\times$10$^{-5}$--0.2$^b$ & \citet{2002OLEB...32..209M}, & Photodestruction & 1.0$\times$10$^{-4}$ photon$^{-1}$ $^c$ & \citet{2015AsBio..15..221P} \\
  & & \citet{2008OLEB...38..383L} &  Seepage & 2.6 mm solution d$^{-1}$ & \citet{Reference116}, \\
 & & & & & \citet{Reference425} \\
 & & & Hydrolysis & 4.8$\times$10$^{-10}$ s$^{-1}$ & \citet{Reference45}
 \\
   & & & & & \\
    Cytosine & 0.036$^d$  & \citet{Yi_et_al2020}, & Photodestruction & 1.0$\times$10$^{-4}$ photon$^{-1}$ $^c$ & \citet{2015AsBio..15..221P} \\
  & & \citet{2008OLEB...38..383L} &  Seepage & 2.6 mm solution d$^{-1}$ & \citet{Reference116}, \\
 & & & & & \citet{Reference425} \\
 & & & Hydrolysis & 1.2$\times$10$^{-8}$ s$^{-1}$ & \citet{Reference45}
 \\
   & & & & & \\
  Uracil & 1.7$\times$10$^{-5}$--0.018$^{bd}$  & \citet{2002OLEB...32..209M}, & Photodestruction & 1.0$\times$10$^{-4}$ photon$^{-1}$ $^c$ & \citet{2015AsBio..15..221P} \\
  & & \citet{Yi_et_al2020}, &  Seepage & 2.6 mm solution d$^{-1}$ & \citet{Reference116}, \\
 & & \citet{2008OLEB...38..383L} & & & \citet{Reference425} \\
 & & & Hydrolysis & 1.4$\times$10$^{-11}$ s$^{-1}$ & \citet{Reference45}
 \\
   & & & & & \\
 Thymine & 0.012$^d$  & \citet{Yi_et_al2020}, & Photodestruction & 1.0$\times$10$^{-4}$ photon$^{-1}$ $^c$ & \citet{2015AsBio..15..221P} \\
  & & \citet{2008OLEB...38..383L} &  Seepage & 2.6 mm solution d$^{-1}$ & \citet{Reference116}, \\
 & &  & & & \citet{Reference425} \\
 & & & Hydrolysis & 2.8$\times$10$^{-12}$ s$^{-1}$ & \citet{Reference45}
 \\
   & & & & & \\
  2-Amino-oxazole & 0.0011 & \citet{Yi_et_al2020}, & Photodestruction & 1.0$\times$10$^{-4}$ photon$^{-1}$ $^c$ & \citet{2015AsBio..15..221P} \\
  & & &  Seepage & 2.6 mm solution d$^{-1}$ & \citet{Reference116}, \\
 & & & & & \citet{Reference425} \\
 & & & Hydrolysis & none$^e$ &  \\
    & & & & & \\
  Ribose & 3.6$\times$10$^{-4}$ $^f$ & \citet{Yi_et_al2020}, & Photodestruction & 1.0$\times$10$^{-4}$ photon$^{-1}$ $^c$ & \citet{2015AsBio..15..221P} \\
  & & \citet{Shapiro_1988} &  Seepage & 2.6 mm solution d$^{-1}$ & \citet{Reference116}, \\
 & & & & & \citet{Reference425} \\
 & & & Hydrolysis & none$^g$ & 
 \\
\hline
\multicolumn{6}{l}{\footnotesize $^a$ Yield range is based on experiments with and without catalysts, e.g., ammonium formate.} \\
\multicolumn{6}{l}{\footnotesize $^b$ Lower yield value is from experiment. Upper value is the theoretical yield.} \\
\multicolumn{6}{l}{\footnotesize $^c$ Photodestruction rate assumed to be similar to that of adenine.} \\
\multicolumn{6}{l}{\footnotesize $^d$ Yield is based on radiolytic experiments for the Kiliani--Fischer synthesis of glyceronitrile \citep{Yi_et_al2020}, for which \ce{H2CO} is initially} \\
\multicolumn{6}{l}{\footnotesize produced from irradiated solutions of HCN and water. We multiply the glyceronitrile yield by 3, given 3 times fewer HCN molecules} \\
\multicolumn{6}{l}{\footnotesize  are required for antecedent \ce{H2CO} synthesis. \ce{H2CO} is then assumed to be the limiting reagent in the theoretical reaction from} \\
\multicolumn{6}{l}{\footnotesize \citet{2008OLEB...38..383L}, for which we assume the 100\% theoretical yield.}  \\
\multicolumn{6}{l}{\footnotesize $^e$ There are no known hydrolysis experiments for 2-aminooxazole; however, it is known to be fairly stable \citep{Szabla_et_al2013}. We assume}  \\
\multicolumn{6}{l}{\footnotesize the other two sinks dominate the loss  of this species.}  \\
\multicolumn{6}{l}{\footnotesize $^f$ Ribose yields from lab experiments of the formose reaction are uncertain; however, 1\% has been suggested.} \\
\multicolumn{6}{l}{\footnotesize $^g$ We assume sufficient borate is present in our ponds to stabilize ribose from hydrolysis \citep{Reference61}.}  
\end{longtable*}

We consider two yields for adenine production from HCN in our pond solutions based on experiments: a lower yield of 0.5\% based on aqueous reactions of HCN \citep{Reference27}, and an upper yield of 18\% based on HCN reactions with more ideal conditions for forming adenine (e.g., solutions containing \ce{NH3}, ammonium formate \citep{2002OLEB...32...99H,ReferenceWaka}.

For guanine and uracil we consider lower yields of 0.0067\% and 0.0017\%, respectively, based on experiments of frozen ammonium cyanide solutions \citep{2002OLEB...32..209M}. We use theoretical yields for the upper bounds. Guanine has a theoretical yield of 20\% based on the theoretical HCN-based reaction equation \ce{5 HCN + H2O -> guanine + H2} \citep{2008OLEB...38..383L}. Uracil has a theoretical yield of 50\%; however, it is based on \ce{H2CO} as a limiting reagent (\ce{2 HCN + 2 H2CO -> uracil + H2} \citep{2008OLEB...38..383L}. Experiments of the Kiliani--Fischer synthesis of glyceronitrile produce \ce{H2CO} as an intermediate from aqueous solutions of HCN \citep{Yi_et_al2020}. Yields of glyceronitirile production are 1.2\%; however, theory suggests 3 intermediate HCN molecules are involved in this reaction. Considering this, we apply a yield of 3.6\% for \ce{H2CO} production from HCN. For uracil, this results in an upper yield of 50\% $\times$ 3.6\% = 1.8\%.

Given the lack of experiments producing cytosine and thymine from aqueous HCN, we only consider the theoretical upper yields for these species base on the reaction equations \ce{3 HCN + H2CO -> cytosine} and \ce{2 HCN + 3 H2CO -> thymine + H2O} \citep{2008OLEB...38..383L}. Again, since \ce{H2CO} is the limiting reagent for these theoretical reactions, we apply the yield of 3.6\% for \ce{H2CO} production from HCN, resulting in yields of 3.6\% and 1.2\% for cytosine and thymine production from HCN, respectively.

For 2-aminooxazole, we consider a yield of 0.11\% based on radiolytic experiments of aqueous solutions of HCN \citep{Yi_et_al2020}. 

Finally, experiments of ribose synthesis from \ce{H2CO} have identified ribose as a product, but yields remain uncertain. \citet{Shapiro_1988} suggests 1\% as an upper bound, therefore we consider this yield and also apply the yield of 3.6\% for \ce{H2CO} production from HCN to obtain an overall yield of 0.036\%.

The sink rates for biomolecule photodestruction, seepage, and hydrolysis are chosen to match those modeled in \citet{Reference425} so that we can directly compare our results with their calculations of biomolecule concentrations in WLPs from meteorites and interplanetary dust particles. Photodestruction and seepage rates are consistent among all biomolecules given the lack of experimental data for each individual biomolecule in our study. Biomolecule photodegredation comparison studies show about order of magnitude differences in photostability between biomolecules \citep{2015AsBio..15..221P,Todd_et_al2019}. Therefore, we expect approximately an order of magnitude additional uncertainty in our biomolecule concentrations during the dry phase, when irradiation is turned on.

\beginappendixB

\section*{Appendix B - Supplementary Results}

\subsection*{Rain-out Rates}

\begin{figure*}[!hbtp]
\centering
\includegraphics[width=\linewidth]{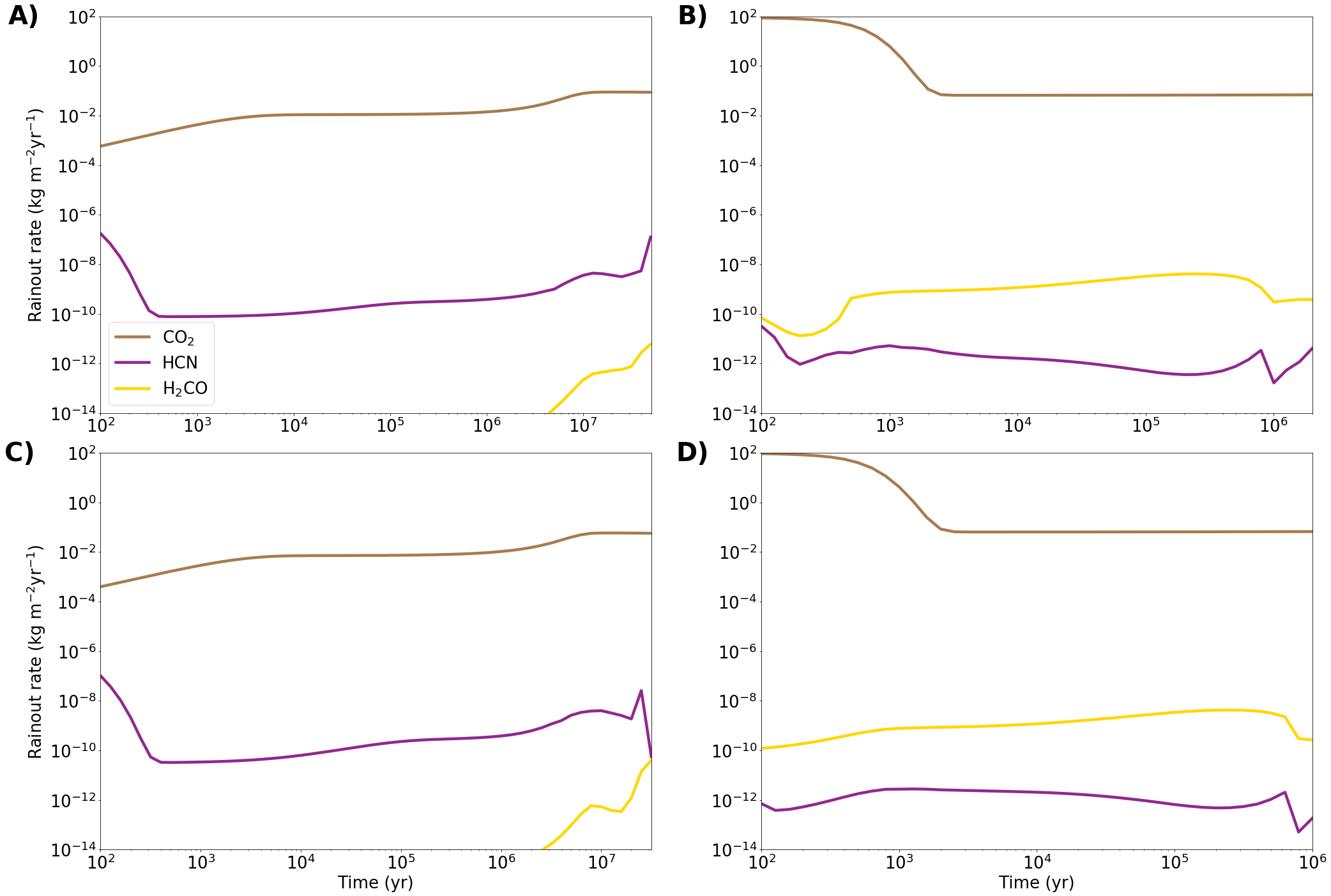}
\caption{Rain-out rates of \ce{HCN}, \ce{CO2}, and \ce{H2CO} from the lowest atmospheric layer as a function of time in our four early Earth models. Models parameters are listed in Table~\ref{models}.}
\label{rain-out_plot}
\end{figure*}

In Figure~\ref{rain-out_plot}, we display the rain-out rates for \ce{HCN}, \ce{H2CO}, and \ce{CO2} as a function of time. These water-soluble species, and a few others, are removed from the lowest layer of our atmospheric models at each time step. The HCN rain-out rate for the early Hadean (reducing) model A is  1.3$\times$10$^{-7}$ kg m$^{-2}$ yr$^{-1}$ at 50 million years. This is about 5 orders of magnitude higher than the HCN rain-out rates for the late Hadean (oxidizing) models of $\sim$10$^{-12}$. The HCN rain-out rate for model C is approximately a factor of 5 lower than model A at 10 million years. We use the 50 million year and 800,000 year HCN rain-out rates from models A and B, respectively, for influx into our subsequent warm little pond models.

The \ce{H2CO} rain-out rate for the early Hadean (reducing) model A is 6.0$\times$10$^{-12}$ kg m$^{-2}$ yr$^{-1}$ at 50 million years. This is $\sim$2 orders of magnitude lower than the \ce{H2CO} rain-out rates for the late Hadean (oxidizing) models of $\sim$10$^{-9}$ kg m$^{-2}$ yr$^{-1}$ at 1 million years.


\subsection*{HCN to CH4 Ratio}

\begin{figure*}[!hbtp]
\centering
\includegraphics[width=\linewidth]{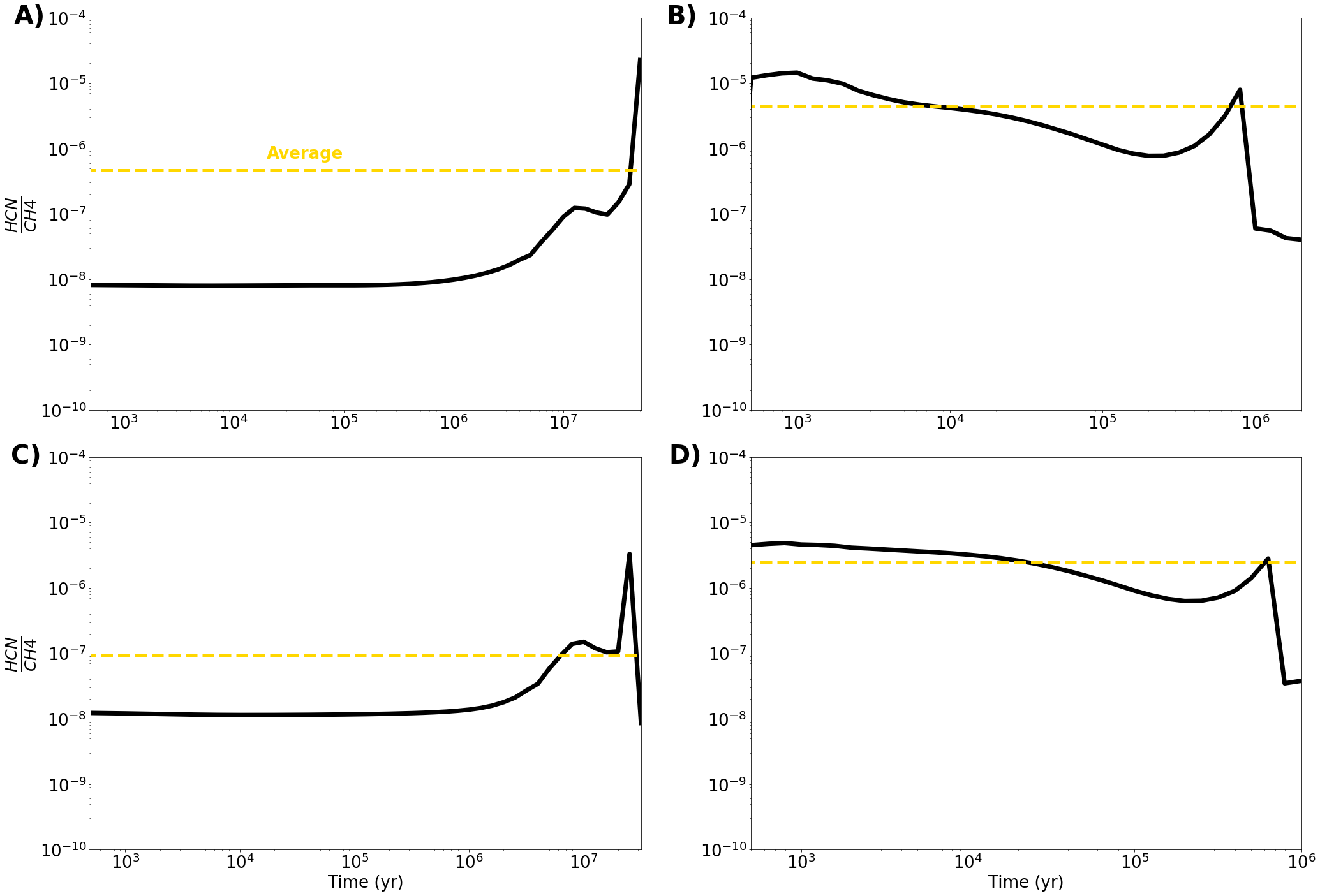}
\caption{The molar abundance ratio of HCN to \ce{CH4} in the lowest atmospheric layer for our four early Earth models. Models parameters are listed in Table~\ref{models}.}
\label{HCN_CH4_plot}
\end{figure*}

In Figure~\ref{HCN_CH4_plot}, we display the molar abundance ratio of HCN/\ce{CH4} in the lowest atmospheric layer from $t$ = 500 years onwards. For the early Hadean (reducing) models A and C, the average HCN/\ce{CH4} ratio during this period is $\sim$1--4$\times$10$^{-7}$. For the late Hadean (oxidizing) models B and D, the average HCN/\ce{CH4} ratios are 3--5$\times$10$^{-6}$. The HCN/\ce{CH4} ratios are constant at 10$^{-8}$ for models A, and C from $t$ = 500 to 1 million years, but vary by up to three orders of magnitude beyond this point. Multiple oxygen species begin to fluctuate in abundance after $\sim$1 million years, including \ce{H2O}, \ce{CO} and \ce{OH}, which affects \ce{HCN} production but not the \ce{CH4} abundance. Similarly, fluctuations in species such as \ce{O2} and \ce{H2} in models B and D after $\sim$600,000 years causes the HCN/\ce{CH4} ratios to fluctuate beyond this point.

\subsection*{No Seepage}

\begin{figure*}[!hbtp]
\centering
\includegraphics[width=\linewidth]{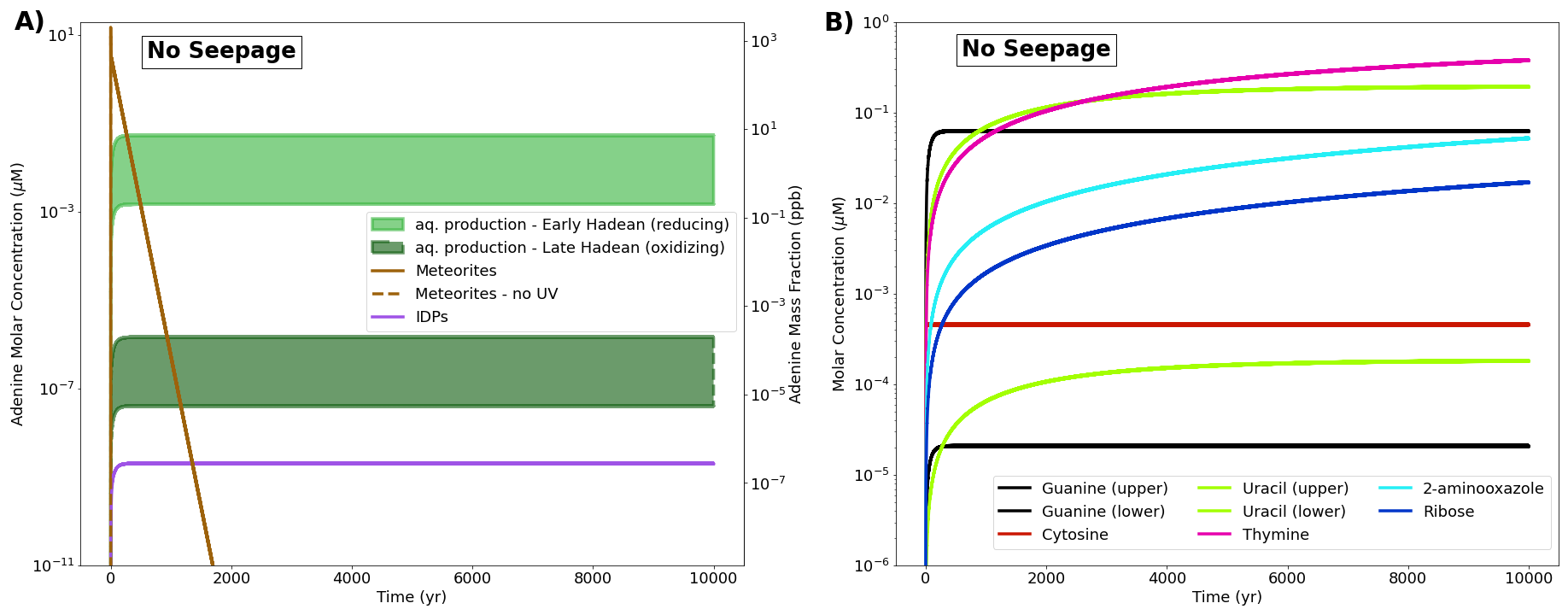}
\caption{The same as Figure~\ref{Pond_Models}, but with seepage turned off.}
\label{NoSeepage}
\end{figure*}

In Figure~\ref{NoSeepage}, we turn off seepage in our model WLP from Figure~\ref{Pond_Models}, and display the results from $t$ = 0 to 10,000 years. This model represents a scenario where the rock pores at the base of the WLP are blocked by, e.g., amphiphilic multilamellar matrices or mineral gels \citep{Damer_Deamer2019,Deamer2017}. In the absence of seepage, hydrolysis takes over as the main biomolecule sink. We see species with high hydrolysis rates such as cytosine, adenine and guanine reach a steady state in just a few hundred years. Hydrolysis rates for uracil and thymine are lower, allowing them to build in concentration to near $\mu$M abundances in 10,000 years. 2-aminooxazole and ribose are not given hydrolysis rates in our models, and thus their concentrations here are to be considered maxima in the absence of aqueous chemical sinks. We summarize the no seepage model biomolecule concentrations at $t$ = 10,000 years in Table~\ref{PeakConcentrations}.

\subsection*{Biomolecule Concentrations}

In Table~\ref{PeakConcentrations}, we display the peak concentrations of biomolecules in our model WLPs for our early Hadean (reducing), late Hadean (oxidizing) and no seepage models.

\begin{table*}[ht!]
\centering
\caption{Summary of the peak concentrations of biomolecules and biomolecule precursors for our fiducial atmosphere and WLP models. A early Hadean (reducing) model A with seepage turned off is also included (see Figure~\ref{rain-out_plot}), representing a pond whose base pores are blocked by, e.g., amphiphilic multilamellar matrices or mineral gels \citep{Damer_Deamer2019,Deamer2017}. \label{PeakConcentrations}} 
\begin{tabular}{lccc}
\\
\multicolumn{1}{c}{Molecule} &  
\multicolumn{1}{l}{Model A (reducing) (nM)} & 
\multicolumn{1}{l}{Model B (oxidizing) (nM)} & 
\multicolumn{1}{l}{Model A (no seepage) (nM)} 
\\[+2mm] \hline \\[-2mm]
HCN & 41.5 & 1.1$\times$10$^{-3}$ & $^a$\\
 \ce{H2CO} (aq. production) & 1.5 & 4.1$\times$10$^{-5}$ & $^a$\\
 \ce{H2CO} (rain-out) & 1.8$\times$10$^{-3}$ & 0.34 & $^a$\\
Adenine & 7.3 & 2.0$\times$10$^{-4}$ & 54.9 \\
 Guanine & 8.2 & 2.2$\times$10$^{-4}$ & 63.3\\
Cytosine & 1.4 & 3.9$\times$10$^{-5}$ & 0.46\\
 Uracil & 0.7 & 2.0$\times$10$^{-5}$ & 195\\
Thymine & 0.5 & 1.3$\times$10$^{-5}$ & 383\\
2-aminooxazole & 0.045 & $\sim$10$^{-6}$ & 52.6\\
Ribose & 0.015 & $\sim$10$^{-7}$ & 17.2\\
\\[-2mm] \hline
\multicolumn{4}{l}{\footnotesize $^a$ We do not estimate hydrolysis rates for these species, which is typically the rate-limiting sink in the absence of seepage.} \\
\end{tabular}
\end{table*}

\subsection*{Maximum Lightning Flash Density}

\begin{figure}[!hbtp]
\centering
\includegraphics[width=\linewidth]{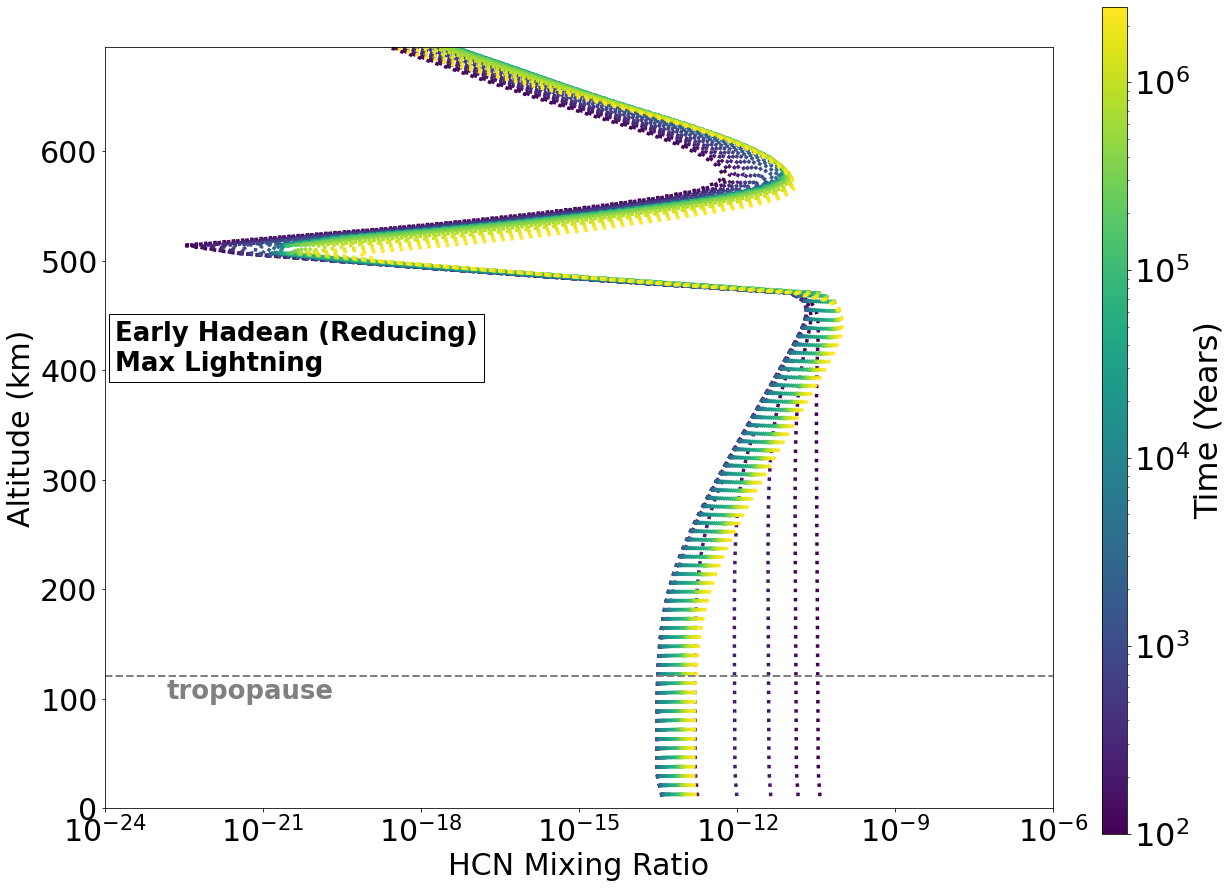}
\caption{The molar abundance of HCN from $t$ = 100 years to 40 million years for model A with an increased lightning flash density of 10$^{4}$ flashes km$^{-2}$ yr$^{-1}$. This flash density is a rough average measured during volcanic eruptions on Earth today, and is 10$^{4}$ times greater than the global average used in our fiducial models \citep{Hodosan2016}. Models parameters are listed in Table~\ref{models}.}
\label{maxlightning_plot}
\end{figure}

In Figure~\ref{maxlightning_plot}, we plot the molar HCN mixing ratio as a function of altitude for our early Hadean (reducing) model A using an increased lightning flash density 10$^4$ times greater than our fiducial rate. This lightning flash density represents an average value measured near volcanic eruptions on Earth today \citep{Hodosan2016}.

We see no changes to the HCN profile when using an enhanced lightning flash density. This suggests that volcanically active regions of early Earth would not have higher HCN abundances than the global average from photochemistry.

\bibliography{Bibliography_Early_Earth}
\bibliographystyle{aasjournal}

\end{document}